\documentclass[aps,prc,onecolumn,floatfix,showpacs,amsmath,amssymb]{revtex4}
\usepackage{graphicx}
\usepackage{amssymb}
\usepackage{dcolumn}
\usepackage{color} 
\begin{document}
\preprint{UCRL-JRNL-230339}
\title{Local three-nucleon interaction from chiral effective field theory}
\author{P. Navr\'atil}
\email[]{navratil1@llnl.gov}
\affiliation{Lawrence Livermore National Laboratory, L-414, P.O. Box 808,
             Livermore, CA  94551, USA}

\date{\today}

\begin{abstract}
The three-nucleon (NNN) interaction derived within the chiral effective field theory
at the next-to-next-to-leading order (N$^2$LO) is regulated with a function
depending on the magnitude of the momentum transfer. The regulated NNN interaction 
is then local in the coordinate space, which is advantages for some many-body techniques.
Matrix elements of the local chiral NNN interaction are evaluated in a three-nucleon
basis. Using the {\it ab initio} no-core shell model (NCSM) the NNN matrix elements
are employed in $^3$H and $^4$He bound-state calculations. 
\end{abstract}
\maketitle
%
\section{\label{sec:intro}Introduction}
Interactions among nucleons are governed by quantum chromodynamics (QCD). 
In the low-energy regime relevant to nuclear structure, 
QCD is non-perturbative, and, therefore, hard to solve. Thus, theory has 
been forced to resort to models for the interaction, which have limited physical basis. 
New theoretical developments, however, allow us connect QCD 
with low-energy nuclear physics. The chiral effective field theory 
($\chi$EFT)~\cite{Weinberg} provides a promising bridge.
Beginning with the pionic or the nucleon-pion system~\cite{bernard95} 
one works consistently with systems of increasing nucleon 
number~\cite{ORK94,Bira,bedaque02a}. 
One makes use of spontaneous breaking of chiral symmetry to systematically 
expand the strong interaction in terms of a generic small momentum
and takes the explicit breaking of chiral symmetry into account by expanding 
in the pion mass. Thereby, the NN interaction, the NNN interaction 
and also $\pi$N scattering are related to each other. 
At the same time, the pion mass dependence of the interaction is known, which will
enable a connection to lattice QCD calculations in the future~\cite{Beane06}.
Nuclear interactions are non-perturbative, because diagrams with purely nucleonic
intermediate states are enhanced \cite{Weinberg}. Therefore, the chiral perturbation 
expansion is performed for the potential (note, however, the discussion in Refs.~\cite{NTK05,Birse,EM06} 
that points out some potential inconsitencies of this approach). Solving the Schr\"odinger equation 
for this potential then automatically sums diagrams with purely nucleonic intermediate 
states to all orders. 
The $\chi$EFT predicts, along with the NN interaction 
at the leading order, an NNN interaction at the 3rd order (next-to-next-to-leading 
order or N$^2$LO)~\cite{Weinberg,vanKolck:1994,Epelbaum:2002}, 
and even an NNNN interaction at the 4th order (N$^3$LO)~\cite{Epelbaum06}.
The details of QCD dynamics are contained in parameters, 
low-energy constants (LECs), not fixed by the symmetry. These parameters 
can be constrained by experiment. At present, high-quality NN potentials 
have been determined at order N$^3$LO~\cite{N3LO}. 
A crucial feature of $\chi$EFT is the consistency between 
the NN, NNN and NNNN parts. As a consequence, at N$^2$LO and N$^3$LO, except 
for two LECs, assigned to two NNN diagrams, 
the potential is fully 
constrained by the parameters defining the NN interaction.

It is of great interest and also a challenge to apply the chiral interactions
in nuclear structure and nuclear reaction calculations.
In a recent work~\cite{Navratil:2007}, the presently available NN potential 
at N$^3$LO~\cite{N3LO} and the NNN interaction at 
N$^2$LO~\cite{vanKolck:1994,Epelbaum:2002} have been applied to the calculation 
of various properties of $s$- and $p$-shell nuclei, using the {\em ab initio} 
no-core shell model (NCSM)~\cite{NCSMC12,NO03}, 
up to now the only approach able to handle the nonlocal $\chi$EFT NN potentials 
for systems beyond $A=4$. In that study, a preferred choice of the two NNN LECs, 
$c_D$ and $c_E$, was found 
and the fundamental importance of the $\chi$EFT NNN interaction was demonstrated 
for reproducing the structure of mid-$p$-shell nuclei. In a subsequent study, the 
same Hamiltonian was used to calculate microscopically the photo-absorption
cross section of $^4$He \cite{Quaglioni:2007}. 

The approach of Ref.~\cite{Navratil:2007} differs in two aspects from 
the first NCSM application of the $\chi$EFT NN+NNN interactions in Ref.~\cite{Nogga06},
which presents a detailed investigation of $^7$Li.
First, a regulator depending on the momentum transfer in the NNN terms was introduced
which results in a local $\chi$EFT NNN interaction. Second, 
the $^4$He binding energy was not used exclusively as the second constraint 
on the $c_D$ and $c_E$ LECs.

A local NNN interaction is advantages for some few- and many-body approaches because
it is simpler to use. At the same time, it is known that details of the NNN interaction
are important for nuclear structure applications. For example, the Urbana IX~\cite{UIX} 
and the Tucson-Melbourne~\cite{TM,TMp,TMprime99} NNN interactions perform differently 
in mid-$p$-shell nuclei~\cite{Pieper:04,GFMC_exc_6_8,NO03} although
their differences appear to be minor. In the Green's function Monte Carlo (GFMC) 
calculations with the AV18 NN potential~\cite{AV18}, the best results 
for $p$-shell nuclei up to $A=10$
are found using the Illinois NNN interaction that augments
the Urbana IX by a two-pion term from the Tucson-Melbourne NNN interaction and by
three-pion terms that in the $\chi$EFT appear beyond 
the N$^3$LO~\cite{GFMC_IL,GFMC_9_10}. Contrary to the Illinois NNN interaction, 
the $\chi$EFT NNN interaction features the above mentioned consistency with the
accompanying NN interaction. Still, interestingly, we found that the nonlocal $\chi$EFT
NNN interaction used in Ref.~\cite{Nogga06} and the local $\chi$EFT NNN interaction 
employed in Ref.~\cite{Navratil:2007} differ to some extent in their description 
of mid-$p$-shell nuclei
with the latter giving results in a better agreement with experiment. Therefore, 
it is important to pay attention to the details of the NNN interaction 
and test different possibilities.

It is the purpose of this paper to elaborate on the details of the local $\chi$EFT NNN
interaction used in Refs.~\cite{Navratil:2007,Quaglioni:2007} and present 
its matrix elements in the three-nucleon basis. Technical details of dealing with NNN
interactions were investigated in many 
papers~\cite{CG81,Friar88,CP93,Huber97,Huber01,Barnea04,Adam04}. 
A new feature in the present work 
is the use of $\chi$EFT contact interactions and a focus on the application within
the {\em ab initio} NCSM. In particular, we demonstrate the binding-energy convergence 
of the three-nucleon and four-nucleon systems with the $\chi$EFT NN+NNN interactions
using the {\em ab initio} NCSM. In Sect.~\ref{sec:local_NNN}, the local $\chi$EFT NNN
interaction is discussed and compared to the nonlocal version of Ref.~\cite{Epelbaum:2002}.
Its three-nucleon matrix elements are given term by term. In Sect.~\ref{sec:H3_He4}, the
$^3$H and $^4$He binding energy and radius calculation results using the N$^3$LO 
$\chi$EFT NN interaction of Ref.~\cite{N3LO} and the local $\chi$EFT NNN interaction 
are given. Conclusions are drawn in Sect.~\ref{sec:Concl}.

\section{\label{sec:local_NNN}Local $\chi$EFT NNN interaction at N$^2$LO}

The NNN interaction appearing at the third order (N$^2$LO) of the $\chi$EFT comprises
of three parts: (i) The two-pion exchange, (ii) the one-pion exchange plus contact
and the three-nucleon contact. In this section, we discuss all the parts in detail
and present the three-nucleon matrix elements of all the terms. For the two parts that
contain the contact interactions, we also discuss in detail the impact of different 
regularization schemes.
 
\subsection{Three-nucleon coordinates}

We use the following definitions of the Jacobi coordinates 
\begin{eqnarray}
\vec{\xi}_1 &=& \frac{1}{\sqrt{2}} (\vec{r}_1 - \vec{r}_2)  \label{Jacobi_r_1} \; ,\\ 
\vec{\xi}_2 &=& \sqrt{\frac{2}{3}} \left(\frac{1}{2}(\vec{r}_1 + \vec{r}_2) 
- \vec{r}_3\right) \label{Jacobi_r_2}  \; , 
\end{eqnarray}
and associated momenta 
\begin{eqnarray}
\vec{\pi}_1 &=& \frac{1}{\sqrt{2}} (\vec{p}_1 - \vec{p}_2) \label{Jacobi_p_1} \; ,\\ 
\vec{\pi}_2 &=& \sqrt{\frac{2}{3}} \left(\frac{1}{2}(\vec{p}_1 + \vec{p}_2) 
- \vec{p}_3\right) \; .  \label{Jacobi_p_2} 
\end{eqnarray}
%

We also define the momenta transferred by nucleon 2 and nucleon 3:
\begin{eqnarray}\label{Q_Qp}
\vec{Q} &=& \vec{p}_2^\prime - \vec{p}_2 = -\frac{1}{\sqrt{2}}(\vec{\pi}_1^\prime-\vec{\pi}_1)
+\frac{1}{\sqrt{6}}(\vec{\pi}_2^\prime-\vec{\pi}_2) \; ,\\ 
\vec{Q}^\prime &=& \vec{p}_3^\prime - \vec{p}_3 = -\sqrt{\frac{2}{3}} 
(\vec{\pi}_2^\prime-\vec{\pi}_2) \; , 
\end{eqnarray}
where the primed coordinates refer to the initial momentum and the unprimed to the final momentum
of the nucleon.

\subsection{General structure of three-nucleon interaction and its matrix element}

The NNN interaction is symmetric under permutation of the three nucleon indexes.
It can be written as a sum of three pieces related by particle permutations:
\begin{equation}\label{W}
W=W_1+W_2+W_3   
\end{equation}
To obtain its matrix element in an antisymmetrized three-nucleon basis we need to consider just
a single term, e.g. $W_1$. In this paper, we use the basis of harmonic oscillator (HO) wave functions.
However, most of the expressions have general validity.
Following notation of Ref.~\cite{TIHO:2000}, a general matrix element 
can be written as
\begin{eqnarray}\label{mat_el_W}
\langle N i J T | W | N' i^\prime J T \rangle = 3 \langle N i J T | W_1 | N^\prime i^\prime J T \rangle
&=& 3 \sum \langle nlsjt, {\cal N L J} || N i J T \rangle 
\langle n^\prime l^\prime s^\prime j^\prime t^\prime, 
{\cal N^\prime L^\prime J^\prime} || N^\prime i^\prime J T \rangle
\nonumber \\
& & 
\times \langle (nlsjt, {\cal N L J})  J T | W_1 | (n^\prime l^\prime s^\prime j^\prime t^\prime, 
{\cal N^\prime L^\prime J^\prime}) J T \rangle   \; ,
\end{eqnarray}
where $| N i J T \rangle$ is an antisymmetrized three-nucleon state with $N=2n+l+2{\cal N}+{\cal L}$,
$i$ an additional quantum number and $J$ and $T$ the total angular momentum and total isospin, respectively.
The parity of the state is $(-1)^N$. The state $|(nlsjt, {\cal N L J})  J T\rangle$ is a product
of the HO wave functions $\langle\vec{\xi}_1|nl\rangle$ and $\langle\vec{\xi}_2|{\cal N L}\rangle$ 
associated with the coordinates~(\ref{Jacobi_r_1}) and (\ref{Jacobi_r_2}), respectively.
This state is antisymmetrized
only with respect to the exchange of nucleons 1 and 2, i.e. $(-1)^{l+s+t}=-1$. The coefficient 
of fractional parentage $\langle nlsjt, {\cal N L J} || N i J T \rangle $ is
calculated according to Ref.~\cite{TIHO:2000}.

\subsection{N$^2$LO three-nucleon interaction contact term}

\begin{figure}[hbtp]
  \includegraphics*[width=0.15\columnwidth]
   {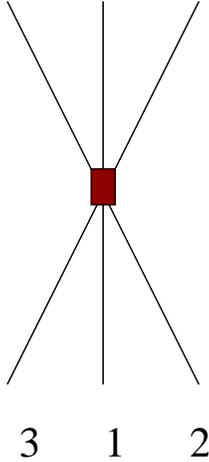}
  \caption{
Contact interaction NNN term of the N$^2$LO $\chi$EFT.
  \label{fig_contact}}
\end{figure}

We start our discussion with the most trivial part of the $\chi$EFT N$^2$LO NNN interaction, 
the three-nucleon contact term
\begin{eqnarray}\label{W1_cont}
W_1^{\rm cont} &=& E \vec{\tau}_2 \cdot \vec{\tau}_3 
\delta(\vec{r}_1-\vec{r}_2) \delta(\vec{r}_3-\vec{r}_1)
=E \vec{\tau}_2 \cdot \vec{\tau}_3 \frac{1}{(2\pi)^6} \frac{1}{(\sqrt{3})^3}
\int {\rm d}\vec{\pi}_1 {\rm d}\vec{\pi}_2 {\rm d}\vec{\pi}_1^\prime {\rm d}\vec{\pi}_2^\prime
| \vec{\pi}_1 \vec{\pi}_2 \rangle \langle \vec{\pi}_1^\prime \vec{\pi}_2^\prime | \; ,
\end{eqnarray}
with $E=\frac{c_E}{F_\pi^4\Lambda_\chi}$ where $\Lambda_\chi$ is the chiral symmetry breaking scale 
of the order of the $\rho$ meson mass and $F_\pi=92.4$~MeV is the weak pion decay constant. 
The $c_E$ is a low-energy constant (LEC) from the chiral Lagrangian of order one.
The corresponding diagram is shown in Fig.~\ref{fig_contact}.
This term was regulated in Ref.~\cite{Epelbaum:2002} by a regulator dependent 
on the sum of Jacobi momenta squared:
\begin{eqnarray}\label{W1_cont_ENGKMW}
W_1^{\rm cont,ENGKMW} &=& E \vec{\tau}_2 \cdot \vec{\tau}_3 \frac{1}{(2\pi)^6} \frac{1}{(\sqrt{3})^3}
\int {\rm d}\vec{\pi}_1 {\rm d}\vec{\pi}_2 {\rm d}\vec{\pi}_1^\prime {\rm d}\vec{\pi}_2^\prime
| \vec{\pi}_1 \vec{\pi}_2 \rangle F(\textstyle{\frac{1}{2}(\pi_1^2+\pi_2^2)};\Lambda) 
F(\textstyle{\frac{1}{2}}(\pi_1^{\prime 2}+\pi_2^{\prime 2});\Lambda)
\langle \vec{\pi}_1^\prime \vec{\pi}_2^\prime | \; ,
\end{eqnarray}
with the regulator function 
\begin{equation}\label{F_q}
F(q^2;\Lambda)= \exp(-q^4/\Lambda^4)   
\end{equation}
with the limit 
$F(q^2;\Lambda\rightarrow \infty)= 1$.   
%
This was in particular convenient as the calculations were performed in momentum space.

Alternatively, let us consider a regulator dependent on momentum transfer:
\begin{eqnarray}\label{W1_cont_mt}
W_1^{\rm cont,Q} &=& E \vec{\tau}_2 \cdot \vec{\tau}_3 \frac{1}{(2\pi)^6} \frac{1}{(\sqrt{3})^3}
\int {\rm d}\vec{\pi}_1 {\rm d}\vec{\pi}_2 {\rm d}\vec{\pi}_1^\prime {\rm d}\vec{\pi}_2^\prime
| \vec{\pi}_1 \vec{\pi}_2 \rangle F(\vec{Q}^2;\Lambda) 
F(\vec{Q}^{\prime 2};\Lambda)
\langle \vec{\pi}_1^\prime \vec{\pi}_2^\prime | 
\nonumber \\
&=& E \vec{\tau}_2 \cdot \vec{\tau}_3 \int {\rm d}\vec{\xi}_1 {\rm d}\vec{\xi}_2 
| \vec{\xi}_1 \vec{\xi}_2 \rangle Z_0(\textstyle{\sqrt{2}}\xi_1;\Lambda) 
Z_0(|\textstyle{\frac{1}{\sqrt{2}}}\vec{\xi}_1+\textstyle{\sqrt{\frac{3}{2}}}\vec{\xi}_2|;\Lambda)
\langle \vec{\xi}_1 \vec{\xi}_2 | \; , 
\end{eqnarray}
where we introduced the function
\begin{equation}\label{Z_0}
Z_0(r;\Lambda)= \frac{1}{2\pi^2} \int {\rm d} q q^2 j_0(qr) F(q^2;\Lambda) \;.   
\end{equation}
This results in an interaction local in coordinate space because of the dependence
of the regulator function on differences of initial and final Jacobi momenta.
An interaction local in coordinate space may be more convenient for some methods.
In fact, most of the NNN interactions used in few-body calculations, such as
the Tucson-Melbourne (TM$^\prime$)~\cite{TMp,TMprime99}, Urbana IX (UIX)~\cite{UIX} 
or Illinois 2 (IL2)~\cite{GFMC_IL}, are local in coordinate space. 

\begin{figure}[hbtp]
  \includegraphics*[width=0.15\columnwidth]
   {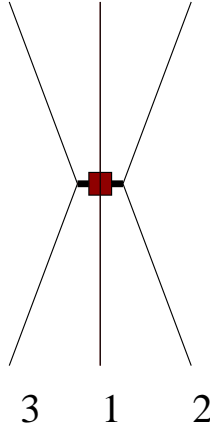}
  \caption{
Contact interaction NNN term of the N$^2$LO $\chi$EFT regulated by a function depending 
on momentum transfer.
  \label{fig_contact_Q}}
\end{figure}

The two alternatively regulated contact interactions lead to different three-nucleon
matrix elements. The interaction~(\ref{W1_cont_ENGKMW}) gives
\begin{eqnarray}\label{mat_el_W1_cont_ENGKMW}
&&\langle (nlsjt, {\cal N L J})  J T | W_1^{\rm cont,ENGKMW} | 
(n^\prime l^\prime s^\prime j^\prime t^\prime, {\cal N^\prime L^\prime J^\prime}) J T \rangle   
\nonumber \\
&=& 
E \frac{1}{2\sqrt{3}\pi^4} \delta_{l0}\delta_{{\cal L}0}\delta_{l^\prime 0}\delta_{{\cal L}^\prime 0}
\delta_{ss^\prime}\delta_{sj}\delta_{s^\prime j^\prime} \delta_{{\cal J}\frac{1}{2}} 
\delta_{{\cal J}^\prime \frac{1}{2}}
\hat{t}\hat{t}^\prime (-1)^{t+t^\prime+T+\textstyle{\frac{1}{2}}}
\left\{ \begin{array}{ccc} t  & t^\prime  & 1 \\
  \textstyle{\frac{1}{2}} & \textstyle{\frac{1}{2}} & \textstyle{\frac{1}{2}}
\end{array}\right\}
\left\{ \begin{array}{ccc} t  & t^\prime  & 1 \\
  \textstyle{\frac{1}{2}} & \textstyle{\frac{1}{2}} & T
\end{array}\right\} 
\nonumber \\
&&
\times \int {\rm d}\pi_1 {\rm d}\pi_2 \pi_1^2 \pi_2^2 (-1)^{(n+{\cal N})} 
R_{n0}(\pi_1,\textstyle{\frac{1}{b}}) R_{{\cal N}0}(\pi_2,\textstyle{\frac{1}{b}}) 
F(\textstyle{\frac{1}{2}(\pi_1^2+\pi_2^2)};\Lambda) 
\nonumber \\
&&
\times \int {\rm d}\pi_1^\prime {\rm d}\pi_2^\prime 
\pi_1^{\prime 2} \pi_2^{\prime 2}
(-1)^{(n^\prime+{\cal N}^\prime)}
R_{n^\prime 0}(\pi^\prime_1,\textstyle{\frac{1}{b}}) 
R_{{\cal N}^\prime 0}(\pi^\prime_2,\textstyle{\frac{1}{b}}) 
F(\textstyle{\frac{1}{2}}(\pi_1^{\prime 2}+\pi_2^{\prime 2});\Lambda) \; ,
\end{eqnarray}
while the interaction~(\ref{W1_cont_mt}) results in the following matrix element:
\begin{eqnarray}\label{mat_el_W1_cont_mt}
&&\langle (nlsjt, {\cal N L J})  J T | W_1^{\rm cont,Q} | 
(n^\prime l^\prime s^\prime j^\prime t^\prime, {\cal N^\prime L^\prime J^\prime}) J T \rangle   
\nonumber \\
&=& 
E 6 \delta_{ss^\prime} 
\hat{t}\hat{t}^\prime (-1)^{t+t^\prime+T+\textstyle{\frac{1}{2}}}
\left\{ \begin{array}{ccc} t  & t^\prime  & 1 \\
  \textstyle{\frac{1}{2}} & \textstyle{\frac{1}{2}} & \textstyle{\frac{1}{2}}
\end{array}\right\}
\left\{ \begin{array}{ccc} t  & t^\prime  & 1 \\
  \textstyle{\frac{1}{2}} & \textstyle{\frac{1}{2}} & T
\end{array}\right\} 
\hat{j}\hat{j}^\prime \hat{\cal J}\hat{\cal J}^\prime \hat{l}^\prime \hat{\cal L}^\prime
(-1)^{J-\textstyle{\frac{1}{2}}+{\cal J}^\prime - {\cal J}+l+{\cal L}+s}
\nonumber \\
&&\times \sum_X (-1)^X \hat{X}^2
\left\{ \begin{array}{ccc} l^\prime & l & X \\
       j & j^\prime & s
\end{array}\right\}
\left\{ \begin{array}{ccc} j & j^\prime & X \\
       {\cal J}^\prime & {\cal J} & J
\end{array}\right\}
\left\{ \begin{array}{ccc} {\cal J}^\prime & {\cal J} & X \\
    {\cal L} & {\cal L}^\prime & \textstyle{\frac{1}{2}} 
\end{array}\right\}
(l^\prime 0 X 0 | l 0) ({\cal L}^\prime 0 X 0| {\cal L} 0 )
\nonumber \\
&&
\times \int {\rm d}\xi_1 {\rm d}\xi_2
\xi_1^{2} \xi_2^{2}
R_{n l}(\xi_1,\textstyle{b}) 
R_{{\cal N} {\cal L}}(\xi_2,\textstyle{b}) 
R_{n^\prime l^\prime}(\xi_1,\textstyle{b}) 
R_{{\cal N}^\prime {\cal L}^\prime}(\xi_2,\textstyle{b}) 
Z_0(\textstyle{\sqrt{2}}\xi_1;\Lambda) 
Z_{0,X}(\sqrt{\textstyle{\frac{1}{2}}}\xi_1,\sqrt{\textstyle{\frac{3}{2}}}\xi_2;\Lambda) \;.
\end{eqnarray}
In the above expressions, we have introduced the radial HO wave functions $R_{nl}$ 
with the oscillator parameter $b$ and, further, a new function
%
\begin{equation}\label{Z_0L}
Z_{0,X}(r_1,r_2;\Lambda)= \frac{1}{2\pi^2} \int {\rm d} q q^2 j_X(qr_1) j_X(qr_2) F(q^2;\Lambda)  \; . 
\end{equation}
We also introduced the customary abriviation $\hat{l}=\sqrt{2l+1}$.
It should be noted that the two differently regulated contact interaction have
different tensorial structure. One would perhaps expect that matrix elements of 
a local interaction will be easier to calculate. This is not the case for the
discussed contact interaction. From Eq.~(\ref{mat_el_W1_cont_ENGKMW}) we can see
that the term (\ref{W1_cont_ENGKMW}) acts only in $S$-waves. On the other hand, the local interaction 
(\ref{W1_cont_mt}) acts in higher partial waves as well as seen from Eq.~(\ref{mat_el_W1_cont_mt}).
We display this schematically in Fig.~\ref{fig_contact_Q} by breaking the symmetry of
the pure contact interaction diagram (Fig.~\ref{fig_contact}) and showing the finite range of 
the momentum transfer regulated interaction.
Using $F(q^2;\Lambda\rightarrow \infty)= 1$, 
$Z_0(r,\Lambda\rightarrow \infty)=\frac{1}{4\pi r^2} \delta(r)$ and 
$Z_{0,X}(r_1,r_2;\Lambda\rightarrow\infty)= \frac{1}{4\pi r_1^2} \delta(r_1-r_2)$,
it is straightforward to verify that in the limit $\Lambda\rightarrow \infty$ both expressions 
(\ref{mat_el_W1_cont_ENGKMW}) and (\ref{mat_el_W1_cont_mt}) lead to the same result. 
For completeness, 
let us note that in Ref.~\cite{Epelbaum:2002} a matrix element of $\vec{\tau}_1 \cdot \vec{\tau}_2$
was calculated, i.e. $\langle \vec{\tau}_1 \cdot \vec{\tau}_2\rangle 
= 6 \delta_{tt^\prime} (-1)^{t-1}
\left\{ \begin{array}{ccc} \textstyle{\frac{1}{2}}  & \textstyle{\frac{1}{2}}  & t \\
  \textstyle{\frac{1}{2}} & \textstyle{\frac{1}{2}} & 1
\end{array}\right\}$, 
instead of $\vec{\tau}_2 \cdot \vec{\tau}_3$ as we do 
in Eq.~(\ref{mat_el_W1_cont_ENGKMW}). Either choice lead to identical matrix element 
in the three-nucleon antisymmetrized basis (\ref{mat_el_W}). This is not the case once we regulate
with the momentum transfer. Our choice in (\ref{W1_cont_mt}) results in the same isospin-coordinate
structure as that obtained in Ref.~\cite{vanKolck:1994}.

\subsection{Transformation of the momentum part of the NNN interaction}\label{mom_transform}

A general NNN interaction term is a product of isospin, spin and momentum parts.
%
In this subsection, we manipulate the momentum part. We only consider the case
of the regulator function $F(q^2,\Lambda)$ depending on transfered momentum. 
The momentum part of a general term $W_1$ can be schematically written as
\begin{equation}\label{TNI_p}
g_{K_1}(|\vec{Q}|;\Lambda) g_{K_2}(|\vec{Q}^\prime|;\Lambda) 
(Y_{K_1}(\hat{Q}) Y_{K_2}(\hat{Q}^\prime))^{(K)}    \; , 
\end{equation}
with $K_1+K_2$ even and with $\vec{Q}$ and $\vec{Q}^\prime$ defined 
by Eqs.~(\ref{Jacobi_p_1}) and (\ref{Jacobi_p_2}), respectively. For coordinates and
momenta, $\hat{Q}$ denotes the angular part of the vector $\vec{Q}$.
A transformation of (\ref{TNI_p}) to coordinate space leads to a local interaction
\begin{eqnarray}\label{TNI_p_r}
&&\frac{1}{(2\pi)^6} \frac{1}{(\sqrt{3})^3}
\int {\rm d}\vec{\pi}_1 {\rm d}\vec{\pi}_2 {\rm d}\vec{\pi}_1^\prime {\rm d}\vec{\pi}_2^\prime
| \vec{\pi}_1 \vec{\pi}_2 \rangle
g_{K_1}(|\vec{Q}|;\Lambda)
g_{K_2}(|\vec{Q}^\prime|;\Lambda)
(Y_{K_1}(\hat{Q}) Y_{K_2}(\hat{Q}^\prime))^{(K)}
\langle \vec{\pi}_1^\prime \vec{\pi}_2^\prime | 
\nonumber\\
&=& i^{K_1+K_2} \int {\rm d}\vec{\xi}_1 {\rm d}\vec{\xi}_2 |\vec{\xi}_1 \vec{\xi}_2 \rangle
f_{K_1}(\sqrt{2}\xi_1;\Lambda) f_{K_2}(|\textstyle{\frac{1}{\sqrt{2}}}\vec{\xi}_1
+\textstyle{\sqrt{\frac{3}{2}}}\vec{\xi}_2|;\Lambda)
(Y_{K_1}(\hat{\xi}_1)
Y_{K_2}(\widehat{\textstyle{\frac{1}{\sqrt{2}}}\vec{\xi}_1
+\textstyle{\sqrt{\frac{3}{2}}}\vec{\xi}_2}))^{(K)}
\langle \vec{\xi}_1 \vec{\xi}_2 | \; .
\end{eqnarray}
Using (\ref{Jacobi_r_1}) and (\ref{Jacobi_r_2}), we note that 
$\sqrt{2}\vec{\xi}_1=\vec{r}_1-\vec{r}_2$ and 
$\textstyle{\frac{1}{\sqrt{2}}}\vec{\xi}_1
+\textstyle{\sqrt{\frac{3}{2}}}\vec{\xi}_2=\vec{r}_1-\vec{r}_3$.
In the above equation, we have introduced a new function using the relation
\begin{equation}
i^K f_K(r;\Lambda) Y_{Kk}(\hat{r}) = \frac{1}{(2\pi)^3}\int {\rm d}\vec{q} 
\;\;{\rm e}^{i\vec{q}\cdot\vec{r}} g_K(q;\Lambda) Y_{Kk}(\hat{q})  \; , 
\end{equation}
which implies
\begin{equation}\label{f_K_gen}
f_K(r;\Lambda)=\frac{1}{2\pi^2} \int {\rm d} q q^2 j_K(qr) g_K(q;\Lambda)   \; .
\end{equation}
We manipulate Eq.~(\ref{TNI_p_r}) first by utilizing the spherical harmonics 
relation
\begin{equation}\label{Ylm_expansion}
Y_{K_2 k_2}(\widehat{\vec{r}_1+\vec{r}_2})=\sum_{K_3=0}^{K_2} 
\frac{\sqrt{4\pi}}{\hat{K}_3} 
\left[\binom{2K_2+1}{2K_3}\right]^{\textstyle{\frac{1}{2}}}
r_1^{K_3} r_2^{K_2-K_3} 
|\vec{r}_1+\vec{r}_2|^{-K_2} 
(Y_{K_3}(\hat{r}_1)Y_{K_2-K_3}(\hat{r}_2))^{K_2}_{k_2} \; ,
\end{equation}
and, second, by the following expansion involving the functions depending on
$|\textstyle{\frac{1}{\sqrt{2}}}\vec{\xi}_1+\textstyle{\sqrt{\frac{3}{2}}}\vec{\xi}_2|$:
\begin{equation}\label{f_K_r1_r2}
f_{K_2}(|\vec{r}_1+\vec{r}_2|;\Lambda) |\vec{r}_1+\vec{r}_2|^{-K_2}
= 4\pi \sum_{X M_X} f_{K_2,X}(r_1,r_2;\Lambda) (-1)^X Y^*_{X M_X}(\hat{r}_1)
Y_{X M_X}(\hat{r}_2) \; ,
\end{equation}
with the function $f_{K_2,X}(r_1,r_2;\Lambda)$ given by
\begin{equation}\label{f_K_X}
f_{K_2,X}(r_1,r_2;\Lambda)= \frac{2}{\pi} \int {\rm d} q q^2 j_X(qr_1) j_X(qr_2) 
\int {\rm d} r r^2 j_0(qr) \frac{f_{K_2}(r;\Lambda)}{r^{K_2}} \; ,
\end{equation}
or, equivalently, by
\begin{equation}\label{f_K_X_Legan}
f_{K_2,X}(r_1,r_2;\Lambda)=\textstyle{\frac{1}{2}} \int_{-1}^1 {\rm d}u P_X(u)
\frac{f_{K_2}(\sqrt{r_1^2+r_2^{2}-2r_1 r_2 u};\Lambda)}{(r_1^2+r_2^{2}-2r_1 r_2 u)^{K_2}} \; .
\end{equation}
Using (\ref{Ylm_expansion}) and (\ref{f_K_r1_r2}), the term (\ref{TNI_p_r}) is re-written in the form
\begin{eqnarray}\label{TNI_p_r_2}
&=& i^{K_1+K_2} \int {\rm d}\vec{\xi}_1 {\rm d}\vec{\xi}_2 |\vec{\xi}_1 \vec{\xi}_2 \rangle
f_{K_1}(\sqrt{2}\xi_1;\Lambda) 
\sum_{K_3=0}^{K_2} \sum_{XYZV} \left[\binom{2K_2+1}{2K_3}\right]^{\textstyle{\frac{1}{2}}}
(\textstyle{\frac{1}{\sqrt{2}}}\xi_1)^{K_3} (\textstyle{\sqrt{\frac{3}{2}}}\xi_2)^{K_2-K_3}
f_{K_2,X}(\textstyle{\frac{1}{\sqrt{2}}}\xi_1,\textstyle{\sqrt{\frac{3}{2}}}\xi_2;\Lambda)
\nonumber
\\
&&\times
\hat{X}^2\widehat{K_2-K_3} \hat{K}_1 \hat{K}_2 \hat{Y} (-1)^{K_1+Y+Z+K}
\left\{ \begin{array}{ccc} K_1 & Y & V \\
       Z & K & K_2
\end{array}\right\} 
\left\{ \begin{array}{ccc} Y & X & K_3 \\
       K_2-K_3 & K_2 & Z
\end{array}\right\}
\nonumber
\\
&&\times 
(X 0 K_3 0 | Y 0) (X 0 K_2-K_3 0 | Z 0) (Y 0 K_1 0 | V 0)
(Y_V(\hat{\xi}_1) Y_Z(\hat{\xi}_2))^{(K)}_k
\langle \vec{\xi}_1 \vec{\xi}_2 | \; ,
\end{eqnarray}
which is convenient for matrix element calculations.

\subsection{One-pion-exchange plus contact N$^2$LO NNN term}

\begin{figure}[hbtp]
  \includegraphics*[width=0.15\columnwidth]
   {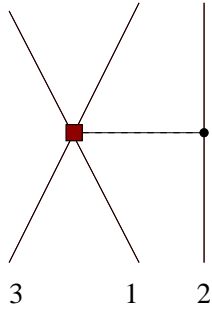}
  \caption{
One-pion exchage plus contact NNN interaction term of the N$^2$LO $\chi$EFT.
  \label{fig_onepion_contact}}
\end{figure}

\begin{figure}[hbtp]
  \includegraphics*[width=0.15\columnwidth]
   {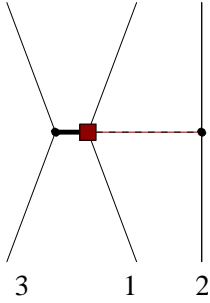}
  \caption{
One-pion exchage plus contact NNN interaction term of the N$^2$LO $\chi$EFT regulated 
by a function depending on momentum transfer.
  \label{fig_onepion_contact_Q}}
\end{figure}

We are now in a position to discuss the one-pion exchange plus contact term that
appears at the N$^2$LO. Following Ref.~\cite{Epelbaum:2002}, we can write the $W_1$ term 
contribution as
\begin{eqnarray}\label{W1_onepi_cont}
W_1^{\rm 1\pi\_cont} &=& -D\frac{1}{(2\pi)^6}\frac{g_{\rm A}}{8F_{\pi}^2}\vec{\tau}_2\cdot\vec{\tau}_3
\left[\frac{1}{\vec{Q}^{\prime 2}+M_\pi^2}\vec{\sigma}_2\cdot\vec{Q}^\prime \vec{\sigma}_3\cdot\vec{Q}^\prime
+\frac{1}{\vec{Q}^{2}+M_\pi^2}\vec{\sigma}_2\cdot\vec{Q} \vec{\sigma}_3\cdot\vec{Q} \right] \; ,
\end{eqnarray}
with $D=\frac{c_D}{F_\pi^2\Lambda_\chi}$, where $c_D$ is a LEC from the chiral Lagrangian of order one.
A diagramatic depiction of the second term in the parenthesis is presented 
in Fig.~\ref{fig_onepion_contact}. The first term corresponds to the exchange of $2\leftrightarrow 3$.
Using the regulator dependent on the sum of Jacobi momenta squared of Ref.~\cite{Epelbaum:2002},
this term can be cast in the form
\begin{eqnarray}\label{W1_onepi_cont_ENGKMW}
W_1^{\rm 1\pi\_cont,{\rm ENGKMW}} &=& - D \frac{1}{(2\pi)^6}\frac{g_{\rm A}}{8F_{\pi}^2} \frac{1}{(\sqrt{3})^3}
\int {\rm d}\vec{\pi}_1 {\rm d}\vec{\pi}_2 {\rm d}\vec{\pi}_1^\prime {\rm d}\vec{\pi}_2^\prime
| \vec{\pi}_1 \vec{\pi}_2 \rangle F(\textstyle{\frac{1}{2}(\pi_1^2+\pi_2^2)};\Lambda) 
\nonumber \\
&&\times
\vec{\tau}_2\cdot\vec{\tau}_3 \left[\frac{1}{\vec{Q}^{\prime 2}+M_\pi^2} 
\vec{\sigma}_2\cdot\vec{Q}^{\prime}\vec{\sigma}_3\cdot\vec{Q}^{\prime}
+\frac{1}{\vec{Q}^{2}+M_\pi^2} 
\vec{\sigma}_2\cdot\vec{Q}\vec{\sigma}_3\cdot\vec{Q}\right]
F(\textstyle{\frac{1}{2}}(\pi_1^{\prime 2}+\pi_2^{\prime 2});\Lambda)
\langle \vec{\pi}_1^\prime \vec{\pi}_2^\prime | \; .
\end{eqnarray}
On the other hand, with a regulator dependent on momentum transfer, we get
\begin{eqnarray}\label{W1_onepi_cont_mt}
W_1^{\rm 1\pi\_cont,{\rm Q}} &=& - D \frac{1}{(2\pi)^6}\frac{g_{\rm A}}{8F_{\pi}^2} \frac{1}{(\sqrt{3})^3}
\int {\rm d}\vec{\pi}_1 {\rm d}\vec{\pi}_2 {\rm d}\vec{\pi}_1^\prime {\rm d}\vec{\pi}_2^\prime
| \vec{\pi}_1 \vec{\pi}_2 \rangle F(\vec{Q}^2;\Lambda) 
\nonumber \\
&&\times
\vec{\tau}_2\cdot\vec{\tau}_3 \left[\frac{1}{\vec{Q}^{\prime 2}+M_\pi^2} 
\vec{\sigma}_2\cdot\vec{Q}^{\prime}\vec{\sigma}_3\cdot\vec{Q}^{\prime}
+\frac{1}{\vec{Q}^{2}+M_\pi^2} 
\vec{\sigma}_2\cdot\vec{Q}\vec{\sigma}_3\cdot\vec{Q}\right]
F(\vec{Q}^{\prime 2};\Lambda)
\langle \vec{\pi}_1^\prime \vec{\pi}_2^\prime | \; ,
\end{eqnarray}
%
which leads to a term local in coordinate space. We depict the second term in the parenthesis
of (\ref{W1_onepi_cont_mt}) schematically in Fig.~\ref{fig_onepion_contact_Q}. 
The first term corresponds to the exchange of $2\leftrightarrow 3$.
This choice of regulation results in 
spin-isopin-coordinate structure that also appears in NNN terms obtained 
in Ref.~\cite{vanKolck:1994}. We note
that a somewhat different spin-isopin structure was used for pion-range-short-range NNN terms
in Refs.~\cite{Huber01} and \cite{Adam04}. In Ref.~\cite{Huber01} in particular, the $\vec{\sigma}$
and $\vec{\tau}$ operators were associated with the active nucleon 1, i.e.
\begin{eqnarray}\label{W1_onepi_cont_mts1}
W_1^{\rm 1\pi\_cont,{\rm Q}_{\sigma_1}} &=& 
- D \frac{1}{(2\pi)^6}\frac{g_{\rm A}}{8F_{\pi}^2} \frac{1}{(\sqrt{3})^3}
\int {\rm d}\vec{\pi}_1 {\rm d}\vec{\pi}_2 {\rm d}\vec{\pi}_1^\prime {\rm d}\vec{\pi}_2^\prime
| \vec{\pi}_1 \vec{\pi}_2 \rangle F(\vec{Q}^2;\Lambda) 
\nonumber \\
&&\times
\left[\vec{\tau}_1\cdot\vec{\tau}_3 \frac{1}{\vec{Q}^{\prime 2}+M_\pi^2} 
\vec{\sigma}_1\cdot\vec{Q}^{\prime}\vec{\sigma}_3\cdot\vec{Q}^{\prime}
+\vec{\tau}_1\cdot\vec{\tau}_2\frac{1}{\vec{Q}^{2}+M_\pi^2} 
\vec{\sigma}_1\cdot\vec{Q}\vec{\sigma}_2\cdot\vec{Q}\right]
F(\vec{Q}^{\prime 2};\Lambda)
\langle \vec{\pi}_1^\prime \vec{\pi}_2^\prime | \; ,
\end{eqnarray}
This change does not alter
the matrix element of (\ref{W1_onepi_cont_ENGKMW}) in the antisymmetrized three-nucleon basis.
It will lead to a difference in the matrix element of (\ref{W1_onepi_cont_mt}).
However, the dependence on the regulator is a higher order
effect than the $\chi$EFT expansion order used to derive the NNN interaction. Therefore,
these differences should have only minor overall effect. In fact, we confirmed in nuclear structure
calculations such as those described in Ref.~\cite{Navratil:2007} that impact of the choice
(\ref{W1_onepi_cont_ENGKMW}) or (\ref{W1_onepi_cont_mt}) is small in particular when 
the natural LECs values are used ($|c_D|\approx 1$). However, a more significant impact 
of the choice 
of the regulator in particular on spin-orbit force sensitive observables is
observed in the case of the two-pion-exchange terms as discussed in the Introduction.

Due to the antisymmetry of the three-nucleon wave functions in (\ref{mat_el_W}), it is sufficient
to consider just one term of the two in parenthesis 
in (\ref{W1_onepi_cont_ENGKMW}) and (\ref{W1_onepi_cont_mt}) and multiply the result by two.
Using the first term, the matrix element of (\ref{W1_onepi_cont_ENGKMW}) with the regulator dependent 
on the sum of Jacobi momenta squared is obtained in the form
\begin{eqnarray}\label{mat_el_W1_onepi_cont_ENGKMW}
&&\langle (nlsjt, {\cal N L J})  J T | W_1^{\rm 1\pi\_cont,{\rm ENGKMW}} | 
(n^\prime l^\prime s^\prime j^\prime t^\prime, {\cal N^\prime L^\prime J^\prime}) J T \rangle   
\nonumber \\
&=&
- D \frac{g_{\rm A}}{F_{\pi}^2} 
\frac{1}{2\sqrt{3}\pi^4} \delta_{l0} \delta_{l^\prime 0} \delta_{sj}\delta_{s^\prime j^\prime}
\hat{t}\hat{t}^\prime 
\left\{ \begin{array}{ccc} t  & t^\prime  & 1 \\
  \textstyle{\frac{1}{2}} & \textstyle{\frac{1}{2}} & \textstyle{\frac{1}{2}}
\end{array}\right\}
\left\{ \begin{array}{ccc} t  & t^\prime  & 1 \\
  \textstyle{\frac{1}{2}} & \textstyle{\frac{1}{2}} & T
\end{array}\right\}
(-1)^{(n+{\cal N}+n^\prime+{\cal N}^\prime)+({\cal L}+{\cal L}^\prime)/2}
\nonumber \\
&&
\times
\hat{j}\hat{j}^\prime \hat{\cal J}\hat{\cal J}^\prime (-1)^{J-{\cal J}+T+\textstyle{\frac{1}{2}}}
\left\{ \begin{array}{ccc} j & j^\prime & 1 \\
  \textstyle{\frac{1}{2}} & \textstyle{\frac{1}{2}} & \textstyle{\frac{1}{2}}
\end{array}\right\}
\left\{ \begin{array}{ccc} j & j^\prime & 1 \\
       {\cal J}^\prime & {\cal J} & J
\end{array}\right\}
\sum_{K=0,2} \hat{K} (1 0 1 0 | K 0) 
\left\{ \begin{array}{ccc} {\cal J} & {\cal L} & \textstyle{\frac{1}{2}} \\
      {\cal J}^\prime & {\cal L}^\prime & \textstyle{\frac{1}{2}} \\
      1 & K & 1 
\end{array}\right\}
\nonumber \\
&&
\times
\sum_{K_1=0}^K \widehat{K-K_1} \left[ \binom{2K+1}{2K_1} \right]^{\textstyle{\frac{1}{2}}}
(-1)^{{\cal L}+K_1} 
\sum_X \hat{X}\hat{\cal L}^\prime (K_1 0 X 0|{\cal L} 0)
({\cal L}^\prime 0 K-K_1 0 | X 0)
\left\{ \begin{array}{ccc} {\cal L}^\prime & K-K_1 & X \\
       K_1 & {\cal L} & K
\end{array}\right\} 
\nonumber \\
&&
\times
\int {\rm d}\pi_1 {\rm d}\pi_2 {\rm d}\pi_1^\prime {\rm d}\pi_2^\prime 
\pi_1^2 \pi_2^2 \pi_1^{\prime 2} \pi_2^{\prime 2}
R_{n0}(\pi_1,\textstyle{\frac{1}{b}}) R_{{\cal N L}}(\pi_2,\textstyle{\frac{1}{b}})  
F(\textstyle{\frac{1}{2}(\pi_1^2+\pi_2^2)};\Lambda) 
\nonumber \\
&&
\times
R_{n^\prime 0}(\pi^\prime_1,\textstyle{\frac{1}{b}}) 
R_{{\cal N}^\prime {\cal L}^\prime}(\pi^\prime_2,\textstyle{\frac{1}{b}}) 
F(\textstyle{\frac{1}{2}}(\pi_1^{\prime 2}+\pi_2^{\prime 2});\Lambda)
\pi_2^{K_1} \pi_2^{\prime K-K_1} g_{K,X}(\pi_2,\pi_2^\prime) \; ,
\end{eqnarray}
where we introduced the function
\begin{equation}\label{g_KX}
g_{K,X}(p,p^\prime) = \frac{2}{\pi} \int {\rm d}q q^2 dr r^2 
\frac{q^{2-K}}{\textstyle{\frac{2}{3}}q^2+M_\pi^2} j_0(qr) j_X(pr) j_X(p^\prime r) \; ,
\end{equation}
which can be alternatively evaluated through
\begin{equation}\label{g_KX_Legan}
g_{K,X}(p,p^\prime) = \textstyle{\frac{1}{2}} \int_{-1}^1 {\rm d}u P_X(u)
\frac{\sqrt{p^2+p^{\prime 2}-2pp^\prime u}^{2-K}}
{\textstyle{\frac{2}{3}}(p^2+p^{\prime 2}-2pp^\prime u)^2+M_\pi^2} \; .
\end{equation}
The matrix element (\ref{mat_el_W1_onepi_cont_ENGKMW}) was first derived 
in Ref.~\cite{Epelbaum:2002}.

For the one-pion exchange plus contact term (\ref{W1_onepi_cont_mt}) 
with the regulator dependent on momentum transfer, we present the matrix element 
obtained using both terms in the parenthesis of (\ref{W1_onepi_cont_mt}).
Due to the three-nucleon wave function antisymmetry, both contributions lead
to the same result for (\ref{mat_el_W}). One can take the advantage of this feature and use the
alternative calculations to check the correctness of the numerical code.
First, we take the first part of (\ref{W1_onepi_cont_mt}) and get
\begin{eqnarray}\label{mat_el_W1_onepi_cont_mt}
&&\langle (nlsjt, {\cal N L J})  J T | W_1^{\rm 1\pi\_cont,{\rm Q}} | 
(n^\prime l^\prime s^\prime j^\prime t^\prime, {\cal N^\prime L^\prime J^\prime}) J T \rangle   
\nonumber \\
&=&
- D \frac{9 g_{\rm A}}{F_{\pi}^2} 
\hat{t}\hat{t}^\prime (-1)^{t+t^\prime+T+\textstyle{\frac{1}{2}}}
\left\{ \begin{array}{ccc} t  & t^\prime  & 1 \\
  \textstyle{\frac{1}{2}} & \textstyle{\frac{1}{2}} & \textstyle{\frac{1}{2}}
\end{array}\right\}
\left\{ \begin{array}{ccc} t  & t^\prime  & 1 \\
  \textstyle{\frac{1}{2}} & \textstyle{\frac{1}{2}} & T
\end{array}\right\}
\hat{j}\hat{j}^\prime \hat{\cal J}\hat{\cal J}^\prime (-1)^{J-{\cal J}+s+j^\prime}
\left\{ \begin{array}{ccc} s & s^\prime & 1 \\
  \textstyle{\frac{1}{2}} & \textstyle{\frac{1}{2}} & \textstyle{\frac{1}{2}}
\end{array}\right\}
\nonumber \\
&&
\times
\sum_{K=0,2} \hat{K} (-1)^{(K/2)} (1 0 1 0 | K 0) 
\sum_{K_2,K_3} \hat{K}_2 \hat{K}_3 \hat{l}^\prime \hat{\cal L}^\prime 
(l^\prime 0 K_2 0 |l 0) ({\cal L}^\prime 0 K_3 0 |{\cal L} 0) 
\nonumber \\
&&
\times
\sum_Z \hat{Z}^2 
\left\{ \begin{array}{ccc} l & s & j \\
             l^\prime & s^\prime & j^\prime \\
      K_2 & 1 & Z 
\end{array}\right\}
\left\{ \begin{array}{ccc} {\cal L} & \textstyle{\frac{1}{2}} & {\cal J}  \\
      {\cal L}^\prime & \textstyle{\frac{1}{2}} & {\cal J}^\prime  \\
      K_3 & 1 & Z 
\end{array}\right\}
\left\{ \begin{array}{ccc} j & j^\prime & Z \\
       {\cal J}^\prime & {\cal J} & J
\end{array}\right\}
\left\{ \begin{array}{ccc} K_2 & 1 & Z \\
       1 & K_3 & K
\end{array}\right\} 
\nonumber \\
&&
\times
\sum_{K_1=0}^K \widehat{K-K_1} 
\left[\binom{2K+1}{2K_1} \right]^{\textstyle{\frac{1}{2}}} 
(-1)^{K_1} 
\sum_X (-1)^X \hat{X}^2 (K_1 0 X 0|K_2 0)
(K-K_1 0 X 0 | K_3 0)
\left\{ \begin{array}{ccc} K_1 & X & K_2 \\
       K_3 & K & K-K_1
\end{array}\right\} 
\nonumber \\
&&
\times \int {\rm d}\xi_1 {\rm d}\xi_2
\xi_1^{2} \xi_2^{2}
R_{n l}(\xi_1,\textstyle{b}) 
R_{{\cal N} {\cal L}}(\xi_2,\textstyle{b}) 
R_{n^\prime l^\prime}(\xi_1,\textstyle{b}) 
R_{{\cal N}^\prime {\cal L}^\prime}(\xi_2,\textstyle{b})
\nonumber \\
&&
\times
(\sqrt{\textstyle{\frac{1}{2}}}\xi_1)^{K_1} 
(\sqrt{\textstyle{\frac{3}{2}}}\xi_2)^{K-K_1}
Z_0(\textstyle{\sqrt{2}}\xi_1;\Lambda) 
f_{K,X}(\sqrt{\textstyle{\frac{1}{2}}}\xi_1,\sqrt{\textstyle{\frac{3}{2}}}\xi_2;\Lambda) \; ,
\end{eqnarray}
with the functions
\begin{equation}\label{f_0L}
f_{0,X}(r_1,r_2;\Lambda)= \frac{1}{2\pi^2} \int {\rm d} q q^2 j_X(qr_1) j_X(qr_2) 
\frac{q^2 F(q^2;\Lambda)}{q^2+M_\pi^2} \; ,  
\end{equation}
and
\begin{equation}\label{f_2L}
f_{2,X}(r_1,r_2;\Lambda)= \frac{1}{4\pi^2} \int_0^\infty {\rm d} q q^2 j_X(qr_1) j_X(qr_2) 
\int_q^\infty {\rm d} k k (k^2-q^2)
\frac{F(k^2;\Lambda)}{k^2+M_\pi^2}   \; , 
\end{equation}
which are special cases of (\ref{f_K_X}). An alternative way of evaluating (\ref{f_2L})
is
\begin{equation}\label{f_2L_Legan}
f_{2,X}(r_1,r_2;\Lambda)=\textstyle{\frac{1}{2}} \int_{-1}^1 {\rm d}u P_X(u)
\frac{f_2(\sqrt{r_1^2+r_2^{2}-2r_1 r_2 u};\Lambda)}{r_1^2+r_2^{2}-2r_1 r_2 u} \; ,
\end{equation}
with
\begin{equation}\label{f_2}
f_2(r;\Lambda)=\frac{1}{2\pi^2} \int {\rm d} q q^2 j_2(qr) \frac{q^2 F(q^2;\Lambda)}{q^2+M_\pi^2}
\; .
\end{equation}
We note that (\ref{f_2L_Legan}) is numerically more efficient than (\ref{f_2L}).

Next, we take the second part of (\ref{W1_onepi_cont_mt}), which results in a
simpler expression for one-pion-exchange plus contact N$^2$LO three-nucleon matrix element
in non-antisymmetrized basis:
%
%
%
%
\begin{eqnarray}\label{mat_el_W1_onepi_cont_mt_simpl}
&&\langle (nlsjt, {\cal N L J})  J T | W_1^{\rm 1\pi\_cont,{\rm Q}} | 
(n^\prime l^\prime s^\prime j^\prime t^\prime, {\cal N^\prime L^\prime J^\prime}) J T \rangle   
\nonumber \\
&=&
- D \frac{9 g_{\rm A}}{F_{\pi}^2} 
\hat{t}\hat{t}^\prime (-1)^{t+t^\prime+T+\textstyle{\frac{1}{2}}}
\left\{ \begin{array}{ccc} t  & t^\prime  & 1 \\
  \textstyle{\frac{1}{2}} & \textstyle{\frac{1}{2}} & \textstyle{\frac{1}{2}}
\end{array}\right\}
\left\{ \begin{array}{ccc} t  & t^\prime  & 1 \\
  \textstyle{\frac{1}{2}} & \textstyle{\frac{1}{2}} & T
\end{array}\right\}
\hat{j}\hat{j}^\prime \hat{\cal J}\hat{\cal J}^\prime \hat{s}\hat{s}^\prime
(-1)^{J-{\cal J}+s+j^\prime}
\left\{ \begin{array}{ccc} s & s^\prime & 1 \\
  \textstyle{\frac{1}{2}} & \textstyle{\frac{1}{2}} & \textstyle{\frac{1}{2}}
\end{array}\right\} \hat{l}^\prime \hat{\cal L}^\prime 
\nonumber \\
&&
\times
\sum_{K=0,2} \hat{K} (-1)^{(K/2)} (1 0 1 0 | K 0) \sum_V (-1)^V \hat{V} 
(V 0 l^\prime 0 | l 0)
\sum_X \hat{X}^2 (X 0 K 0 | V 0) (X 0 {\cal L}^\prime 0 | {\cal L} 0)
\nonumber \\
&&
\times
\sum_Z \hat{Z}^2 
\left\{ \begin{array}{ccc} l & s & j \\
             l^\prime & s^\prime & j^\prime \\
      V & 1 & Z 
\end{array}\right\}
\left\{ \begin{array}{ccc} {\cal L} & \textstyle{\frac{1}{2}} & {\cal J}  \\
      {\cal L}^\prime & \textstyle{\frac{1}{2}} & {\cal J}^\prime  \\
      X & 1 & Z 
\end{array}\right\}
\left\{ \begin{array}{ccc} j & j^\prime & Z \\
       {\cal J}^\prime & {\cal J} & J
\end{array}\right\}
\left\{ \begin{array}{ccc} V & 1 & Z \\
       1 & X & K
\end{array}\right\} 
\nonumber \\
&&
\times
\int {\rm d}\xi_1 {\rm d}\xi_2
\xi_1^{2} \xi_2^{2}
R_{n l}(\xi_1,\textstyle{b}) 
R_{{\cal N} {\cal L}}(\xi_2,\textstyle{b}) 
R_{n^\prime l^\prime}(\xi_1,\textstyle{b}) 
R_{{\cal N}^\prime {\cal L}^\prime}(\xi_2,\textstyle{b})
f_K(\textstyle{\sqrt{2}}\xi_1;\Lambda) 
Z_{0,X}(\sqrt{\textstyle{\frac{1}{2}}}\xi_1,\sqrt{\textstyle{\frac{3}{2}}}\xi_2;\Lambda) \; ,
\end{eqnarray}
with
\begin{equation}\label{f_K}
f_K(r;\Lambda)=\frac{1}{2\pi^2} \int {\rm d} q q^2 j_K(qr) \frac{q^2 F(q^2;\Lambda)}{q^2+M_\pi^2}
\; ,
\end{equation}
which is a special case of (\ref{f_K_gen})
and $Z_{0,X}(r_1,r_2;\Lambda)$ given by Eq.~(\ref{Z_0L}).
Both (\ref{mat_el_W1_onepi_cont_mt}) and (\ref{mat_el_W1_onepi_cont_mt_simpl}) are already multiplied
by two in anticipation of the three-nucleon antisymmetry in the final matrix element (\ref{mat_el_W}).

For completeness, we also present the matrix element of the second part of 
(\ref{W1_onepi_cont_mts1}):
\begin{eqnarray}\label{mat_el_W1_onepi_cont_mts1_simpl}
&&\langle (nlsjt, {\cal N L J})  J T | W_1^{\rm 1\pi\_cont,{\rm Q}_{\sigma_1}} | 
(n^\prime l^\prime s^\prime j^\prime t^\prime, {\cal N^\prime L^\prime J^\prime}) J T \rangle   
\nonumber \\
&=&
- D \frac{9 g_{\rm A}}{F_{\pi}^2} 
\delta_{tt^\prime}(-1)^{t-1}
\left\{ \begin{array}{ccc} \textstyle{\frac{1}{2}}  & \textstyle{\frac{1}{2}}  & t \\
  \textstyle{\frac{1}{2}} & \textstyle{\frac{1}{2}} & 1
\end{array}\right\}
\hat{j}\hat{j}^\prime \hat{\cal J}\hat{\cal J}^\prime \hat{s}\hat{s}^\prime
\hat{l}^\prime \hat{\cal L}^\prime 
(-1)^{s+s^\prime+j^\prime+J+{\cal L}+{\cal J}+{\cal J}^\prime+\textstyle{\frac{1}{2}}}
\nonumber \\
&&
\times
\sum_{K=0,2} \hat{K} (-1)^{(K/2)} (1 0 1 0 | K 0) 
\left\{ \begin{array}{ccc} s & s^\prime & K \\
            \textstyle{\frac{1}{2}} & \textstyle{\frac{1}{2}} & 1 \\
      \textstyle{\frac{1}{2}} & \textstyle{\frac{1}{2}} & 1 
\end{array}\right\}
\sum_V \hat{V} (V 0 l^\prime 0 | l 0)
\sum_X \hat{X}^2 (X 0 K 0 | V 0) (X 0 {\cal L}^\prime 0 | {\cal L} 0)
\nonumber \\
&&
\times
\left\{ \begin{array}{ccc} l & s & j \\
             l^\prime & s^\prime & j^\prime \\
      V & K & X 
\end{array}\right\}
\left\{ \begin{array}{ccc} {\cal J} & X & {\cal J}^\prime \\
      j^\prime  & J & j 
\end{array}\right\}
\left\{ \begin{array}{ccc} {\cal J} & X & {\cal J}^\prime \\
      {\cal L}^\prime  & \textstyle{\frac{1}{2}} & {\cal L} 
\end{array}\right\}
\nonumber \\
&&
\times
\int {\rm d}\xi_1 {\rm d}\xi_2
\xi_1^{2} \xi_2^{2}
R_{n l}(\xi_1,\textstyle{b}) 
R_{{\cal N} {\cal L}}(\xi_2,\textstyle{b}) 
R_{n^\prime l^\prime}(\xi_1,\textstyle{b}) 
R_{{\cal N}^\prime {\cal L}^\prime}(\xi_2,\textstyle{b})
f_K(\textstyle{\sqrt{2}}\xi_1;\Lambda) 
Z_{0,X}(\sqrt{\textstyle{\frac{1}{2}}}\xi_1,\sqrt{\textstyle{\frac{3}{2}}}\xi_2;\Lambda) \; ,
\end{eqnarray}
which is still simpler than (\ref{mat_el_W1_onepi_cont_mt_simpl}). 
Again, this matrix element is already multiplied by two in anticipation 
of the three-nucleon antisymmetry in the final matrix element (\ref{mat_el_W}).
The matrix element of the (two-times the) first part of (\ref{W1_onepi_cont_mts1}) 
is given by (\ref{mat_el_W1_onepi_cont_mt}) multiplied by $(-1)^{t+t^\prime+s+s^\prime}$.
Due to the three-nucleon wave function antisymmetry, this contribution leads
to the same result for (\ref{mat_el_W}) as does (\ref{mat_el_W1_onepi_cont_mt_simpl}),
which can be taken advantage of in testing the correctness of numerical calculations.

By comparing (\ref{W1_onepi_cont_ENGKMW}) with (\ref{mat_el_W1_onepi_cont_mt}) 
(or equivalently with (\ref{mat_el_W1_onepi_cont_mt_simpl}) and also with 
(\ref{mat_el_W1_onepi_cont_mts1_simpl}))
we note the different tensorial structure of the matrix elements. When the regulator dependent
on the sum of Jacobi momenta squared is used, only the $l=0$, $l^\prime=0$ partial waves
contribute. This is not the case, when the regulator depending on momentum transfer is
utilized. At the same time, however, we note that in the limit $\Lambda\rightarrow \infty$ 
both expressions (\ref{W1_onepi_cont_ENGKMW}) and (\ref{mat_el_W1_onepi_cont_mt}) as well as
(\ref{mat_el_W1_onepi_cont_mt_simpl}) and (\ref{mat_el_W1_onepi_cont_mts1_simpl}) 
lead to the same result.

\subsection{Two-pion exchange N$^2$LO NNN terms}

\begin{figure}[hbtp]
  \includegraphics*[width=0.2\columnwidth]
   {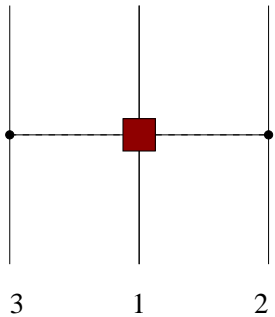}
  \caption{
Two-pion exchange NNN interaction term of the N$^2$LO $\chi$EFT.
  \label{fig_twopion}}
\end{figure}

In this subsection, we present matrix elements of two-pion exchange N$^2$LO NNN terms.
Their schematic depiction is shown in Fig.~\ref{fig_twopion}.
There are three distinct terms associated with three LECs, $c_1$, $c_3$ and $c_4$,
from the chiral Lagrangian, which also appear in the subleading two-pion exchange in the NN
potential. Consequently, values of these LECs expected to be of order one
are typically fixed at the NN level unlike 
the case of the previously introduced $c_D$ (\ref{W1_onepi_cont}) and $c_E$ (\ref{W1_cont}) 
LECs whose values needs to be fixed in systems of more than two nucleons. In the present paper,
we derive only the matrix elements of the two-pion exchange NNN interaction terms regulated
by a function depending on momentum transfer, i.e. terms that are local in coordinate space. 

Following Ref.~\cite{Epelbaum:2002}, the $W_1$ part of the $c_1$ term with the momentum transfer 
regulators can be written as
\begin{eqnarray}\label{W1_twopi_c1}
W_1^{\rm 2\pi\_c1} &=& -c_1 \frac{1}{(2\pi)^6} \frac{4M_\pi^2}{F_\pi^2} 
\frac{g_{\rm A}^2}{4F_{\pi}^2} \vec{\tau}_2\cdot\vec{\tau}_3
F(\vec{Q}^2;\Lambda) \frac{1}{\vec{Q}^{2}+M_\pi^2}
 \vec{\sigma}_2\cdot\vec{Q} \vec{\sigma}_3\cdot\vec{Q}^\prime
\frac{1}{\vec{Q}^{\prime 2}+M_\pi^2} F(\vec{Q}^{\prime 2};\Lambda) \; .
\end{eqnarray}
Using results of Subsect.~\ref{mom_transform}, we find for the $c_1$-term matrix element:
\begin{eqnarray}\label{mat_el_c1}
&&\langle (nlsjt, {\cal N L J})  J T | W_1^{\rm 2\pi\_c1,{\rm Q}} | 
(n^\prime l^\prime s^\prime j^\prime t^\prime, {\cal N^\prime L^\prime J^\prime}) J T \rangle   
\nonumber \\
&=& -c_1 \frac{36 M_\pi^2}{F_\pi^2} \frac{g_{\rm A}^2}{F_{\pi}^2}
\hat{t}\hat{t}^\prime (-1)^{t+t^\prime+T+\textstyle{\frac{1}{2}}}
\left\{ \begin{array}{ccc} t  & t^\prime  & 1 \\
  \textstyle{\frac{1}{2}} & \textstyle{\frac{1}{2}} & \textstyle{\frac{1}{2}}
\end{array}\right\}
\left\{ \begin{array}{ccc} t  & t^\prime  & 1 \\
  \textstyle{\frac{1}{2}} & \textstyle{\frac{1}{2}} & T
\end{array}\right\}
\hat{j}\hat{j}^\prime \hat{\cal J}\hat{\cal J}^\prime \hat{s}\hat{s}^\prime
(-1)^{J-{\cal J}+s+j^\prime}
\left\{ \begin{array}{ccc} s & s^\prime & 1 \\
  \textstyle{\frac{1}{2}} & \textstyle{\frac{1}{2}} & \textstyle{\frac{1}{2}}
\end{array}\right\} \hat{l}^\prime \hat{\cal L}^\prime 
\nonumber \\
&&
\times
\sum_{VR}\hat{V} \hat{R} (V 0 l^\prime 0 | l 0) (R 0 {\cal L}^\prime 0 | {\cal L} 0)
\sum_Y (-1)^{Y}\hat{Y} (Y 0 1 0 | V 0)
\left\{ \begin{array}{ccc} l & s & j \\
             l^\prime & s^\prime & j^\prime \\
      V & 1 & Y
\end{array}\right\}
\left\{ \begin{array}{ccc} {\cal L} & \textstyle{\frac{1}{2}} & {\cal J}  \\
      {\cal L}^\prime & \textstyle{\frac{1}{2}} & {\cal J}^\prime  \\
      R & 1 & Y 
\end{array}\right\}
\left\{ \begin{array}{ccc} j & j^\prime & Y \\
       {\cal J}^\prime & {\cal J} & J
\end{array}\right\}
\nonumber \\
&&
\times
\sum_{K_3=0}^{1} \left[\binom{3}{2K_3}\right]^{\textstyle{\frac{1}{2}}}
\widehat{1-K_3}
\sum_X \hat{X}^2 (X 0 K_3 0|Y 0) (X 0 1-K_3 0 |R 0)
\left\{ \begin{array}{ccc} Y & X & K_3 \\
       1-K_3 & 1 & R
\end{array}\right\} 
\nonumber \\
&&
\times
\int {\rm d}\xi_1 {\rm d}\xi_2
\xi_1^{2} \xi_2^{2}
R_{n l}(\xi_1,\textstyle{b}) 
R_{{\cal N} {\cal L}}(\xi_2,\textstyle{b}) 
R_{n^\prime l^\prime}(\xi_1,\textstyle{b}) 
R_{{\cal N}^\prime {\cal L}^\prime}(\xi_2,\textstyle{b})
\nonumber \\
&&
\times
(\sqrt{\textstyle{\frac{1}{2}}}\xi_1)^{K_3} 
(\sqrt{\textstyle{\frac{3}{2}}}\xi_2)^{1-K_3}
f_1(\textstyle{\sqrt{2}}\xi_1;\Lambda) 
f_{1,X}(\sqrt{\textstyle{\frac{1}{2}}}\xi_1,
\sqrt{\textstyle{\frac{3}{2}}}\xi_2;\Lambda) \; .
\end{eqnarray}
Here we introduced the functions
\begin{equation}\label{f_1}
f_{1}(r;\Lambda)= \frac{1}{2\pi^2} \int {\rm d} q q^2 j_1(q) 
\frac{q F(q^2;\Lambda)}{q^2+M_\pi^2}   \; , 
\end{equation}
and
\begin{equation}\label{f_1L}
f_{1,X}(r_1,r_2;\Lambda)= \frac{1}{2\pi^2} \int_0^\infty {\rm d} q q^2 j_X(qr_1) j_X(qr_2) 
\int_q^\infty {\rm d} k
\frac{k F(k^2;\Lambda)}{k^2+M_\pi^2}  \; ,  
\end{equation}
which are the explicit versions of functions given in Eqs.~(\ref{f_K_gen}) 
and (\ref{f_K_X}), respectively. The function (\ref{f_1L}) can be alternatively
evaluated with the help of the Legendre polynomial:
\begin{equation}\label{f_1L_Legan}
f_{1,X}(r_1,r_2;\Lambda)=\textstyle{\frac{1}{2}} \int_{-1}^1 {\rm d}u P_X(u)
\frac{f_1(\sqrt{r_1^2+r_2^{2}-2r_1 r_2 u};\Lambda)}{\sqrt{r_1^2+r_2^{2}-2r_1 r_2 u}} \; .
\end{equation}

The $W_1$ part of the two-pion exchange $c_3$ term is given by~\cite{Epelbaum:2002}
\begin{eqnarray}\label{W1_twopi_c3}
W_1^{\rm 2\pi\_c3} &=& c_3 \frac{1}{(2\pi)^6} \frac{2}{F_\pi^2} 
\frac{g_{\rm A}^2}{4F_{\pi}^2} \vec{\tau}_2\cdot\vec{\tau}_3
F(\vec{Q}^2;\Lambda) \frac{1}{\vec{Q}^{2}+M_\pi^2}
 \vec{\sigma}_2\cdot\vec{Q} \vec{\sigma}_3\cdot\vec{Q}^\prime
\vec{Q}\cdot\vec{Q}^\prime
\frac{1}{\vec{Q}^{\prime 2}+M_\pi^2} F(\vec{Q}^{\prime 2};\Lambda) \;.
\end{eqnarray}
For its matrix element we find
\begin{eqnarray}\label{mat_el_c3}
&&\langle (nlsjt, {\cal N L J})  J T | W_1^{\rm 2\pi\_c3,{\rm Q}} | 
(n^\prime l^\prime s^\prime j^\prime t^\prime, {\cal N^\prime L^\prime J^\prime}) J T \rangle   
\nonumber \\
&=& c_3 \frac{18}{F_\pi^2} \frac{g_{\rm A}^2}{F_{\pi}^2}
\hat{t}\hat{t}^\prime (-1)^{t+t^\prime+T+\textstyle{\frac{1}{2}}}
\left\{ \begin{array}{ccc} t  & t^\prime  & 1 \\
  \textstyle{\frac{1}{2}} & \textstyle{\frac{1}{2}} & \textstyle{\frac{1}{2}}
\end{array}\right\}
\left\{ \begin{array}{ccc} t  & t^\prime  & 1 \\
  \textstyle{\frac{1}{2}} & \textstyle{\frac{1}{2}} & T
\end{array}\right\}
\hat{j}\hat{j}^\prime \hat{\cal J}\hat{\cal J}^\prime \hat{s}\hat{s}^\prime
(-1)^{J-{\cal J}+s+j^\prime}
\left\{ \begin{array}{ccc} s & s^\prime & 1 \\
  \textstyle{\frac{1}{2}} & \textstyle{\frac{1}{2}} & \textstyle{\frac{1}{2}}
\end{array}\right\} \hat{l}^\prime \hat{\cal L}^\prime 
\nonumber \\
&&
\times
\sum_{K_1 K_2} (-1)^{(K_1+K_2)/2} \hat{K}_1 \hat{K}_2 (1 0 1 0|K_1 0)
(1 0 1 0|K_2 0)
\nonumber \\
&&
\times
\sum_{VR} (-1)^{V+R}\hat{V} \hat{R} (V 0 l^\prime 0 | l 0) 
(R 0 {\cal L}^\prime 0 | {\cal L} 0)
\sum_Y \hat{Y} (Y 0 K_1 0 | V 0)
\nonumber \\
&&
\times
\sum_Z \hat{Z}^2 (-1)^Z 
\left\{ \begin{array}{ccc} l & s & j \\
             l^\prime & s^\prime & j^\prime \\
      V & 1 & Z 
\end{array}\right\}
\left\{ \begin{array}{ccc} {\cal L} & \textstyle{\frac{1}{2}} & {\cal J}  \\
      {\cal L}^\prime & \textstyle{\frac{1}{2}} & {\cal J}^\prime  \\
      R & 1 & Z 
\end{array}\right\}
\left\{ \begin{array}{ccc} j & j^\prime & Z \\
       {\cal J}^\prime & {\cal J} & J
\end{array}\right\}
\left\{ \begin{array}{ccc} Z & 1 & Y \\
       K_1 & V & 1
\end{array}\right\} 
\left\{ \begin{array}{ccc} Z & 1 & Y \\
       K_2 & R & 1
\end{array}\right\} 
\nonumber \\
&&
\times
\sum_{K_3=0}^{K2} \left[\binom{2K_2+1}{2K_3}\right]^{\textstyle{\frac{1}{2}}}
\widehat{K_2-K_3}
\sum_X \hat{X}^2 (X 0 K_3 0|Y 0) (X 0 K_2-K_3 0 |R 0)
\left\{ \begin{array}{ccc} Y & X & K_3 \\
       K_2-K_3 & K_2 & R
\end{array}\right\} 
\nonumber \\
&&
\times
\int {\rm d}\xi_1 {\rm d}\xi_2
\xi_1^{2} \xi_2^{2}
R_{n l}(\xi_1,\textstyle{b}) 
R_{{\cal N} {\cal L}}(\xi_2,\textstyle{b}) 
R_{n^\prime l^\prime}(\xi_1,\textstyle{b}) 
R_{{\cal N}^\prime {\cal L}^\prime}(\xi_2,\textstyle{b})
\nonumber \\
&&
\times
(\sqrt{\textstyle{\frac{1}{2}}}\xi_1)^{K_3} 
(\sqrt{\textstyle{\frac{3}{2}}}\xi_2)^{K_2-K_3}
f_{K_1}(\textstyle{\sqrt{2}}\xi_1;\Lambda) 
f_{K_2,X}(\sqrt{\textstyle{\frac{1}{2}}}\xi_1,
\sqrt{\textstyle{\frac{3}{2}}}\xi_2;\Lambda) \; ,
\end{eqnarray}
with the function $f_{K_1}(r;\Lambda)$ given by (\ref{f_K}),
the function $f_{0,X}(r_1,r_2;\Lambda)$ given by (\ref{f_0L}) and
the function $f_{2,X}(r_1,r_2;\Lambda)$ given by (\ref{f_2L}).

Finally, the $W_1$ part of two-pion exchange $c_4$ term is given by~\cite{Epelbaum:2002}
\begin{eqnarray}\label{W1_twopi_c4}
W_1^{\rm 2\pi\_c4} &=& c_4 \frac{1}{(2\pi)^6} \frac{1}{F_\pi^2} 
\frac{g_{\rm A}^2}{4F_{\pi}^2} \vec{\tau}_1\cdot(\vec{\tau}_2\times\vec{\tau}_3)
F(\vec{Q}^2;\Lambda) \frac{1}{\vec{Q}^{2}+M_\pi^2}
 \vec{\sigma}_2\cdot\vec{Q} \vec{\sigma}_3\cdot\vec{Q}^\prime
\vec{\sigma}_1\cdot(\vec{Q}\times\vec{Q}^\prime)
\frac{1}{\vec{Q}^{\prime 2}+M_\pi^2} F(\vec{Q}^{\prime 2};\Lambda) \; ,
\end{eqnarray}
and for its  matrix element we find
\begin{eqnarray}\label{mat_el_c4}
&&\langle (nlsjt, {\cal N L J})  J T | W_1^{\rm 2\pi\_c4,{\rm Q}} | 
(n^\prime l^\prime s^\prime j^\prime t^\prime, {\cal N^\prime L^\prime J^\prime}) J T \rangle   
\nonumber \\
&=& -c_4 \frac{36^2}{F_\pi^2} \frac{g_{\rm A}^2}{4F_{\pi}^2} 
\hat{t}\hat{t}^\prime (-1)^{T+t^\prime+\textstyle{\frac{1}{2}}}
\left\{ \begin{array}{ccc} t  & t^\prime  & 1 \\
  \textstyle{\frac{1}{2}} & \textstyle{\frac{1}{2}} & T
\end{array}\right\}
\left\{ \begin{array}{ccc} 
  \textstyle{\frac{1}{2}}  & \textstyle{\frac{1}{2}}  & t^\prime \\
  \textstyle{\frac{1}{2}} & \textstyle{\frac{1}{2}} & t \\
   1 & 1 & 1
\end{array}\right\}
\hat{j}\hat{j}^\prime \hat{\cal J}\hat{\cal J}^\prime \hat{s}\hat{s}^\prime
(-1)^{J+{\cal J}+j^\prime}
\hat{l}^\prime \hat{\cal L}^\prime 
\nonumber \\
&&
\times
\sum_{K_1 K_2} (-1)^{(K_1+K_2)/2} \hat{K}_1 \hat{K}_2 (1 0 1 0|K_1 0)
(1 0 1 0|K_2 0)
\nonumber \\
&&
\times
\sum_{VR} (-1)^{V+R}\hat{V} \hat{R} (V 0 l^\prime 0 | l 0) 
(R 0 {\cal L}^\prime 0 | {\cal L} 0)
\sum_Y \hat{Y} (Y 0 K_1 0 | V 0)
\nonumber \\
&&
\times
\sum_Z \hat{Z}^2 (-1)^Z 
\left\{ \begin{array}{ccc} {\cal L} & \textstyle{\frac{1}{2}} & {\cal J}  \\
      {\cal L}^\prime & \textstyle{\frac{1}{2}} & {\cal J}^\prime  \\
      R & 1 & Z 
\end{array}\right\}
\left\{ \begin{array}{ccc} j & j^\prime & Z \\
       {\cal J}^\prime & {\cal J} & J
\end{array}\right\}
\left\{ \begin{array}{ccc} Z & 1 & Y \\
       K_2 & R & 1
\end{array}\right\} 
\nonumber \\
&&
\times
\sum_{K_4} \hat{K}_4^2
\left\{ \begin{array}{ccc} l & s & j \\
             l^\prime & s^\prime & j^\prime \\
      V & K_4 & Z 
\end{array}\right\}
\left\{ \begin{array}{ccc} 
  \textstyle{\frac{1}{2}}  & \textstyle{\frac{1}{2}}  & s \\
  \textstyle{\frac{1}{2}} & \textstyle{\frac{1}{2}} & s^\prime \\
   1 & 1 & K_4
\end{array}\right\}
\left\{ \begin{array}{ccc} Z & 1 & Y \\
       K_1 & V & K_4
\end{array}\right\} 
\left\{ \begin{array}{ccc} 1 & 1 & 1 \\
       1 & K_4 & K_1
\end{array}\right\} 
\nonumber \\
&&
\times
\sum_{K_3=0}^{K2} \left[\binom{2K_2+1}{2K_3}\right]^{\textstyle{\frac{1}{2}}}
\widehat{K_2-K_3}
\sum_X \hat{X}^2 (X 0 K_3 0|Y 0) (X 0 K_2-K_3 0 |R 0)
\left\{ \begin{array}{ccc} Y & X & K_3 \\
       K_2-K_3 & K_2 & R
\end{array}\right\} 
\nonumber \\
&&
\times
\int {\rm d}\xi_1 {\rm d}\xi_2
\xi_1^{2} \xi_2^{2}
R_{n l}(\xi_1,\textstyle{b}) 
R_{{\cal N} {\cal L}}(\xi_2,\textstyle{b}) 
R_{n^\prime l^\prime}(\xi_1,\textstyle{b}) 
R_{{\cal N}^\prime {\cal L}^\prime}(\xi_2,\textstyle{b})
\nonumber \\
&&
\times
(\sqrt{\textstyle{\frac{1}{2}}}\xi_1)^{K_3} 
(\sqrt{\textstyle{\frac{3}{2}}}\xi_2)^{K_2-K_3}
f_{K_1}(\textstyle{\sqrt{2}}\xi_1;\Lambda) 
f_{K_2,X}(\sqrt{\textstyle{\frac{1}{2}}}\xi_1,
\sqrt{\textstyle{\frac{3}{2}}}\xi_2;\Lambda) \; .
\end{eqnarray}
The same functions (\ref{f_0L}), (\ref{f_2L}) and (\ref{f_K}) that were introduced in the $c_3$
term enter the $c_4$ term as well.
%
%
%
%

We note that the local two-pion-exchange terms appear also in the Tucson-Melbourne
NNN interaction~\cite{TM}. The analogous terms to $c_1$, $c_3$ and $c_4$ are present
in particular in the TM$^\prime$ interaction~\cite{TMp,TMprime99}. The TM$^\prime$ parameters
are denoted by $a^\prime$, $b$ and $d$ with the relation to the above $c_1$, $c_3$ and $c_4$
given by 
\begin{eqnarray}\label{TMp_param}
a^\prime&=&\frac{4M_\pi^2}{F_\pi^2}c_1 \; , \\
b&=&\frac{2}{F_\pi^2}c_3 \; , \\
d&=&\frac{-1}{F_\pi^2}c_4   \; . 
\end{eqnarray}
Further, the regulator function $F(q^2;\Lambda)$ is chosen in the form
\begin{equation}
F^{TM}(q^2;\Lambda)=\frac{\Lambda^2-M_\pi^2}{\Lambda^2+q^2} \; .
\end{equation}
This choice allows to evaluate integrals that define the functions $f_K$ analytically.
The analytic expressions can be found, e.g. in Ref.~\cite{Friar88}. In that paper,
the following function is introduced:
\begin{equation}\label{Z_1}
Z_1(r;\Lambda)=\frac{1}{2\pi^2} \int {\rm d} q q^2 j_0(qr) \frac{F(q^2;\Lambda)}{q^2+M_\pi^2} \; .
\end{equation}
Using the properties of spherical Bessel functions, we can easily find relations
between derivatives of $Z_1$ and our $f_K$ functions:
\begin{eqnarray}\label{tilde_Z_1}
f_0(r,\Lambda)&=&-\left(Z_1^{\prime\prime}(r;\Lambda)+\frac{2}{r}Z_1^{\prime}(r;\Lambda)\right) \; ,\\
f_1(r,\Lambda)&=&-Z_1^{\prime}(r;\Lambda) \; \\
f_2(r,\Lambda)&=& 
Z_1^{\prime\prime}(r;\Lambda)-\frac{1}{r}Z_1^{\prime}(r;\Lambda) \; .
\end{eqnarray}
For completeness, we note that a still different notation was used in Ref.~\cite{vanKolck:1994},
where functions $I_{k,l}(r;\Lambda)$ were introduced. They are related to the $Z_0(r,\Lambda)$ 
that we introduced in Eq. (\ref{Z_0}) and to the above $Z_1$ function (\ref{Z_1}) is as follows:
\begin{eqnarray}\label{_I_k_l}
Z_0(r,\Lambda)&=&I_{0,0}(r;\Lambda) \; ,\\
Z_1(r,\Lambda)&=&I_{2,0}(r,\Lambda) \; .
\end{eqnarray}

In Ref.~\cite{NCSM_TM}, the Tucson-Melbourne NNN interaction matrix elements
in the HO basis were calculated using a different algorithm than the one 
used in this paper. That algorithm relied on a completeness relation 
and transformations of HO states
with the help of HO brackets. Even though the algorithm of Ref.~\cite{NCSM_TM} required
calculation of one-dimensional radial integrals, while the the present algorithm requires
evaluation of two-dimensional radial integrals, the present algorithm is substantially
more efficient.

\section{\label{sec:H3_He4}Convergence test for $^3$H and $^4$He}
In this section, we apply the matrix elements of the N$^2$LO $\chi$EFT NNN interaction
obtained in this paper to the NCSM calculation of $^3$H ad $^4$He ground state properties.
As a test of correctness of the computer code, we verified that the new more efficient 
algorithm reproduces the results obtained using the algorithm of Ref.~\cite{NCSM_TM}
for the two-pion-exchange term matrix elements. For the contact terms, 
we verified that in the limit of $\Lambda\rightarrow\infty$, the matrix elements
(\ref{mat_el_W1_cont_mt}) and (\ref{mat_el_W1_cont_ENGKMW}) lead to the same result
and the same is true for matrix elements (\ref{mat_el_W1_onepi_cont_mt_simpl}) 
and (\ref{mat_el_W1_onepi_cont_ENGKMW}). In addition, we benchmarked the computer code
for evaluation of (\ref{mat_el_W1_cont_ENGKMW}) and (\ref{mat_el_W1_onepi_cont_ENGKMW})
with the computer code written by A. Nogga~\cite{Nogga_pc}. Finally, we tested numerically
that the use of (\ref{mat_el_W1_onepi_cont_mt}) results in the same matrix
element as the use of (\ref{mat_el_W1_onepi_cont_mt_simpl}) in the three-nucleon antisymmetrized
basis $| N i J T \rangle$ introduced in Eq.~(\ref{mat_el_W}). The same checks were also
performed for the alternative version of the one-pion-exchange plus contact term ($D$-term)
given in Eq. (\ref{W1_onepi_cont_mts1}). That is, we verified numerically 
that the matrix element (\ref{mat_el_W1_onepi_cont_mts1_simpl}) leads to the same result as  
(\ref{mat_el_W1_onepi_cont_ENGKMW}) in the limit of $\Lambda\rightarrow\infty$ and
the use of (\ref{mat_el_W1_onepi_cont_mts1_simpl}) results in the same matrix
element in the three-nucleon antisymmetrized basis $| N i J T \rangle$ 
as the use of (\ref{mat_el_W1_onepi_cont_mt}) multiplied by $(-1)^{t+t^\prime+s+s^\prime}$.

We use the N$^3$LO NN interaction of Ref.~\cite{N3LO}. We adopt the $c_1$, $c_3$ and $c_4$ LECs
values as well as the value of $\Lambda$ from the N$^3$LO NN interaction of Ref.~\cite{N3LO}
for our local chiral EFT N$^2$LO NNN interaction. The regulator function was chosen in a form
consistent with that used in Refs.~\cite{Epelbaum:2002} and ~\cite{N3LO}: 
$F(q^2;\Lambda)={\rm exp}(-q^4/\Lambda^4)$ (\ref{F_q}).
Values of the $c_D$ and $c_E$ LECs
are constrained by a fit to the $A=3$ system binding energy \cite{Nogga06,Navratil:2007}.
Obviously, additional constraints are needed to uniquely determine values
of $c_D$ and $c_E$, see Refs.~\cite{Epelbaum:2002,Nogga06,Navratil:2007,Bira_cD} 
for discussions of different possibilities.
Here we are interested only in convergence properties of our calculations. Therefore,
we simply select a reasonable value, e.g. $c_D=1$, and follow Ref.~\cite{Navratil:2007} 
and adopt $c_E$ value
as an average of fits to $^3$H and $^3$He binding energies. 
In Table~\ref{tab:param}, we summarize the NNN interaction parameters used in calculations 
described in this section. We note that $^4$He results obtained with the identical Hamiltonian
but with $c_D=-1$ and $c_E=-0.331$ are presented in Ref.~\cite{Quaglioni:2007}.
\begin{table}[hbtp]
  \caption{NNN interaction parameters used in the present calculations. The regulator 
function was chosen in the form $F(q^2;\Lambda)={\rm exp}(-q^4/\Lambda^4)$.
  \label{tab:param}}
  \begin{ruledtabular}
    \begin{tabular}{cccccccccc}
$c_1$ [GeV$^-1$] & $c_3$ [GeV$^-1$] & $c_4$ [GeV$^-1$] & $c_D$ & $c_E$ & 
$\Lambda$ [MeV] & $\Lambda_\chi$ [MeV] & $M_{\pi} [MeV]$ & $g_A$ & $F_{\pi} [MeV]$ \\
\hline
-0.81 & -3.2  & 5.4  & 1.0   & -0.029& 500 & 700 & 138 & 1.29 & 92.4 \\
    \end{tabular}
  \end{ruledtabular}
\end{table}

\begin{figure}[hbtp]
  \includegraphics*[width=0.9\columnwidth]
   {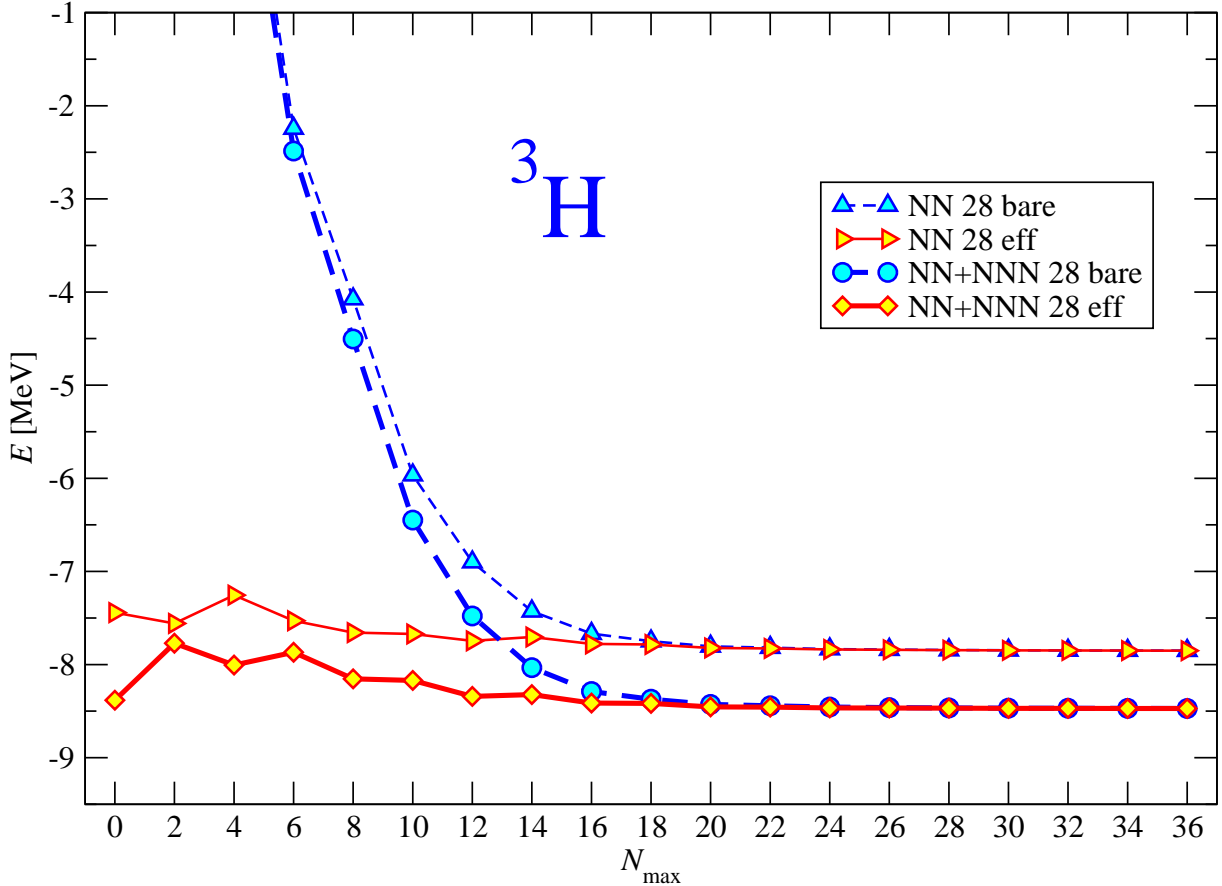}
  \caption{(Color online) $^3$H ground-state energy dependence on the size of the basis.
The HO frequency of $\hbar\Omega=28$ MeV was employed. Results with (thick lines)
and without (thin lines) the NNN interaction are shown. The full lines correspond 
to calculations with two-body effective interaction derived from the chiral NN interaction, 
the dashed lines to calculations with the bare chiral NN interaction. For further details 
see the text.
  \label{gs_H3}}
\end{figure}

\begin{figure}[hbtp]
  \includegraphics*[width=0.9\columnwidth]
   {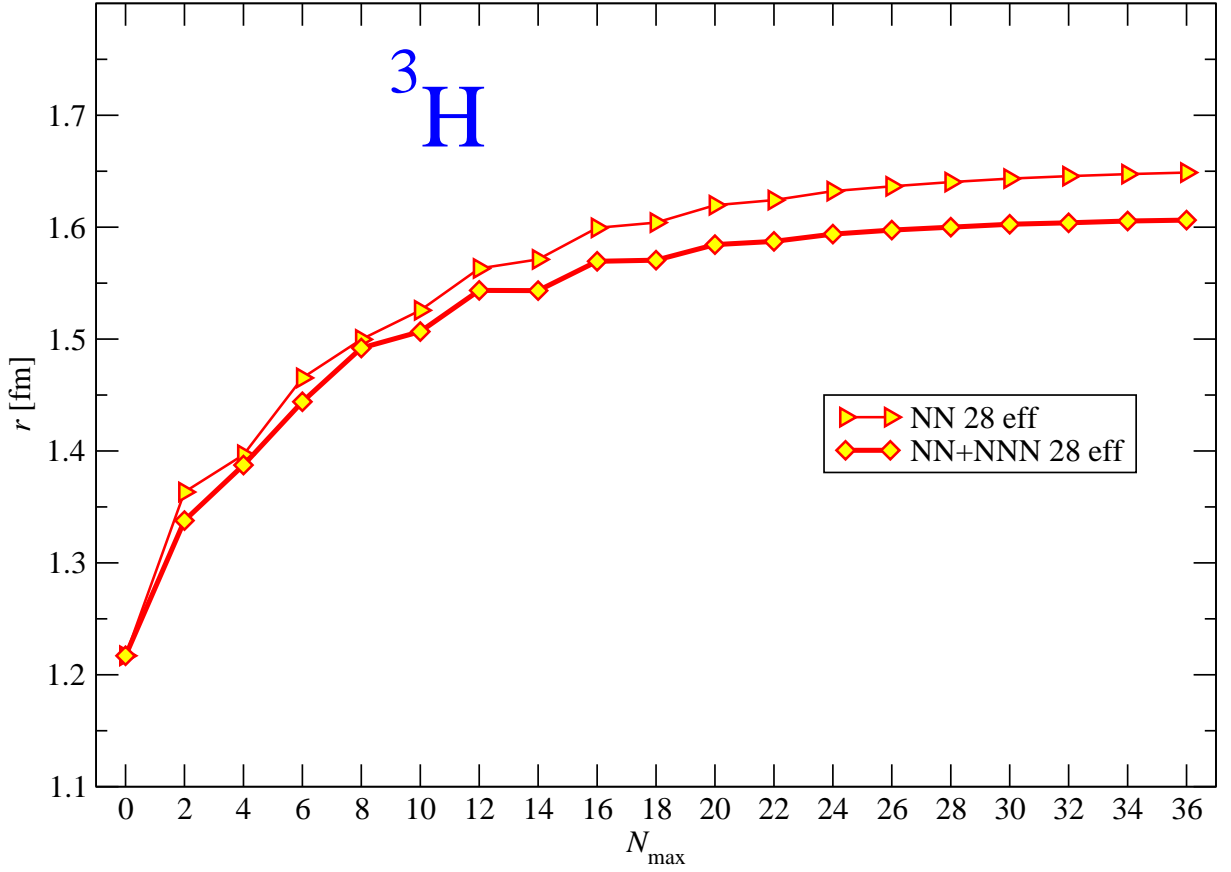}
  \caption{(Color online) $^3$H point-proton rms radius dependence on the size of the basis. 
The HO frequency of $\hbar\Omega=28$ MeV was employed. Results with (thick line)
and without (thin line) the NNN interaction are shown. The two-body effective interaction 
derived from the chiral NN interaction was used in the calculation. For further details 
see the text.
  \label{rp_H3}}
\end{figure}

We use the Jacobi coordinate HO basis antisymmentrized according to the method 
described in Ref.~\cite{TIHO:2000}. In Figs.~\ref{gs_H3} and \ref{rp_H3}, we show the
convergence of the $^3$H ground-state energy and point-proton rms radius, respectively, 
with the size of the basis. Thin lines correspond to results obtained with 
the NN interaction only. Thick lines correspond to calculations that also include the 
NNN interaction. The full lines correspond to calculations with two-body effective
interaction derived from the chiral EFT N$^3$LO NN interaction. The dashed lines correspond 
to calculations with the bare chiral EFT N$^3$LO NN interaction. 
The bare NNN interaction is added to either the bare NN or to the effective NN interaction
in calculations depicted by thick lines. We observe that the convergence is faster when
the two-body effective interaction is used. However, starting at about $N_{\rm max}=24$
the convergence is reached also in calculations with the bare NN interaction. The rate
of convergence also depends on the choice of the HO frequency. In general, it is
always advantageous to use the effective interaction in order to improve the convergence
rate. The $^3$H ground-state energy and point-proton radius results are summarized
in Table~\ref{tab:H3_He4_gs_rp}. The contributions of different NNN terms to the $^3$H 
ground-state energy are presented in Table~\ref{tab:H3_gs}. In addition to results
obtained using the $c_D=1$, we also show in Table~\ref{tab:H3_gs} 
results obtained using $c_D=-1$ and a corresponding 
$c_E$ constrained by the avarage of the $^3$H and $^3$He binding energy fit. 
For completeness, we show results obtained by the two alternative 
one-pion-exchange plus contact terms (\ref{W1_onepi_cont_mt}) and (\ref{W1_onepi_cont_mts1}).
In all cases, the contact $E$-term gives a positive contribution. 
The contribution from the $D$-term changes sign depending on the choice of $c_D$.
Still, the two-pion exchange $c$-terms dominate the NNN expectation value. 

\begin{figure}[hbtp]
  \includegraphics*[width=0.9\columnwidth]
   {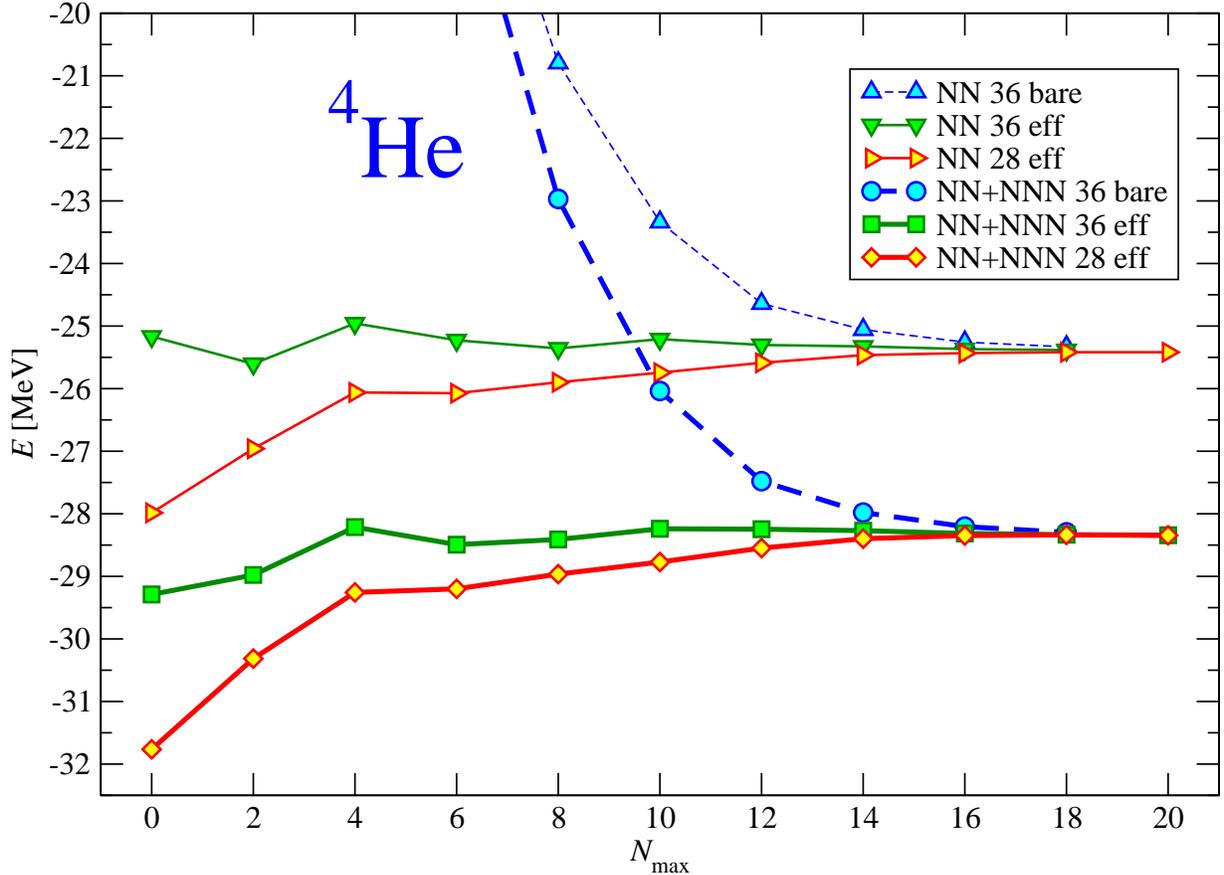}
  \caption{(Color online) $^4$He ground-state energy dependence on the size of the basis.
The HO frequencies of $\hbar\Omega=28$ and 36 MeV was employed. Results with (thick lines)
and without (thin lines) the NNN interaction are shown. The full lines correspond 
to calculations with three-body effective interaction, 
the dashed lines to calculations with the bare interaction. For further details 
see the text.
  \label{gs_He4}}
\end{figure}

\begin{figure}[hbtp]
  \includegraphics*[width=0.9\columnwidth]
   {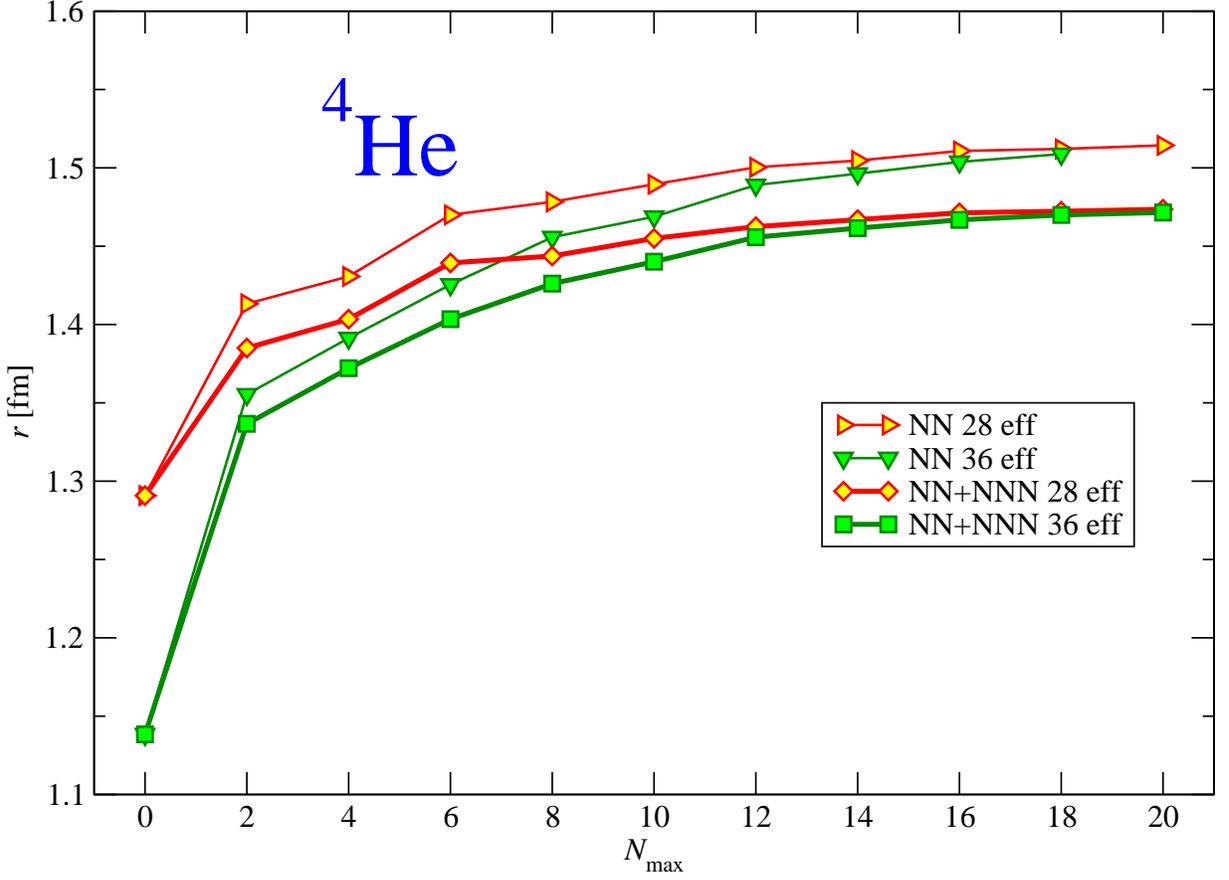}
  \caption{(Color online) $^4$He point-proton rms radius dependence on the size of the basis. 
The HO frequencies of $\hbar\Omega=28$ and 36 MeV was employed. Results with (thick line)
and without (thin line) the NNN interaction are shown. The three-body effective interaction 
was used in the calculation. For further details see the text.
  \label{rp_He4}}
\end{figure}

In Figs.~\ref{gs_He4} and \ref{rp_He4}, we show convergence of the $^4$He ground-state
energy and point-proton rms radius, respectively. The NCSM calculations are perforemed 
in basis spaces up to $N_{\rm max}=20$. Thin lines correspond to results obtained with 
the NN interaction only, while thick lines correspond to calculations that also include the 
NNN interaction. The dashed lines correspond to results obtained with bare interactions.
The full lines correspond to results obtained using three-body effective interaction
(the NCSM three-body cluster approximation). It is apparent that the use of the three-body
effective interaction improves the convergence rate dramatically. We can see that at 
about $N_{\rm max}=18$ the bare interaction calculation reaches convergence as well. 
It should be noted, however, that $p$-shell calculations with the NNN interactions
are presently feasible in model spaces up to $N_{\rm max}=6$ or $N_{\max}=8$. 
The use of the three-body effective interaction is then essential in the $p$-shell 
calculations.

\begin{table}[hbtp]
  \caption{Ground-state energy and point-proton rms radius 
of $^3$H and $^4$He calculated 
using the chiral N$^3$LO NN potential \protect\cite{N3LO} with and without
the local chiral N$^2$LO NNN interaction. The LECs values and other parameters
are given in Table~\protect\ref{tab:param}. The calculations were performed within the 
{\it ab initio} NCSM.
  \label{tab:H3_He4_gs_rp}}
  \begin{ruledtabular}
    \begin{tabular}{cccc}
\multicolumn{4}{c}{$^3$H} \\
                   & NN        &  NN+NNN   & Expt.  \\
\hline
$E_{\rm gs}$ [MeV] & -7.852(5) & -8.473(5) & -8.482  \\
$r_p$ [fm]         &  1.650(5) &  1.608(5) &         \\

\multicolumn{4}{c}{$^4$He} \\
                   & NN         &  NN+NNN     & Expt.    \\
\hline
$E_{\rm gs}$ [MeV] & -25.39(1)  &  -28.34(2)  & -28.296    \\
$r_p$ [fm]         &   1.515(2) &    1.475(2) &   1.455(7) \\ 
    \end{tabular}
  \end{ruledtabular}
\end{table}
\begin{table}[hbtp]
  \caption{Contributions of different NNN terms to the $^3$H ground-state energy.
The $c_D$ and $c_E$ LECs are explicitly shown. Other parameters are given in 
Table~\protect\ref{tab:param}. All energies are given in MeV. The two alternative
one-pion-exchange plus contact terms (\ref{W1_onepi_cont_mt}) and (\ref{W1_onepi_cont_mts1})
are considered.
\label{tab:H3_gs}}
  \begin{ruledtabular}
    \begin{tabular}{cccccc}
\multicolumn{4}{c}{$^3$H} \\
$c_D$  & $c_E$  &  $E_{\rm gs}$ & $c$ terms & $D$ term  & $E$ term \\
\hline
 1.0 (Eq.~(\ref{W1_onepi_cont_mt}))&  -0.029 & -8.473 & -1.01 & 0.13 & 0.03 \\
-1.0 (Eq.~(\ref{W1_onepi_cont_mt}))&  -0.331 & -8.474 & -1.07 &-0.16 & 0.32 \\
 1.0 (Eq.~(\ref{W1_onepi_cont_mts1}))&  -0.159 & -8.471 & -0.99 & 0.005& 0.14 \\
-1.0 (Eq.~(\ref{W1_onepi_cont_mts1}))&  -0.213 & -8.474 & -1.10 &-0.05 & 0.21 \\
    \end{tabular}
  \end{ruledtabular}
\end{table}

We note that NCSM calculations in the three-body cluster approximation are rather
involved. 
The $^4$He NCSM calculation with the three-body effective interaction proceeds 
in three steps. First, we diagonalize the Hamiltonian 
with and without the NNN interaction in a three-nucleon basis for 
all relevant three-body channels. In the second step, we use the three-body solutions 
from the first step to derive three-body effective interactions with and without 
the NNN interaction. By subtracting the two effective interactions
we isolate the NN and NNN contributions. This is needed due to a different 
scaling with particle number of the two- and the three-body interactions. 
The $^4$He efffective interaction is then obtained by adding the two contributions 
with the appropriate scaling factors~\cite{NO03}. In the third step, we diagonalize
the resulting Hamiltonian in the antisymmetrized four-nucleon Jacobi-coordinate HO basis 
to obtain the $^4$He $J^\pi T=0^+ 0$ ground state.
Obviously, in calculations without the NNN interaction, the above three steps are simplified
as no NNN contribution needs to be isolated. In addition, in the case of no NNN interaction,
we may use just the two-body effective interaction (two-body cluster approximation), which
is much simpler. The convergence is slower, however, see discussion in Ref.~\cite{NO02}.
We also note that $^4$He properties with the chiral N$^3$LO NN interaction that we employ
here were calculated using two-body cluster approximation in Ref.~\cite{NC04} and
present results are in agreement with results found there.

Our $^4$He results are summarized in Table~\ref{tab:H3_He4_gs_rp}. We note that
the present NCSM $^3$H and $^4$He results obtained with the chiral N$^3$LO
NN interaction are in a perfect agreement with results obtained using the variational
calculations in the hyperspherical harmonics basis as well as with the Faddeev-Yakubovsky
calculations published in Ref.~\cite{HHnonloc}. A satisfying feature of the present
NCSM calculation is the fact that the rate of convergence is not affected
in any significant way by inclusion of the NNN interaction.

\section{\label{sec:Concl}Conclusions}
In this paper, we regulated the NNN interaction derived within 
the chiral effective field theory at the N$^2$LO 
with a function depending on the magnitude of the momentum transfer. 
The regulated NNN interaction is local in the coordinate space. This is advantages 
for some many-body techniques. In addition, it was found that this interaction
performs sligthtly better in mid-$p$-shell nuclei than its nonlocal counterpart 
\cite{Navratil:2007,Nogga_pc}.
We calculated matrix elements of the local chiral NNN interaction in the 
three-nucleon HO basis and performed calculations for $^3$H and $^4$He 
within the {\em ab initio} NCSM. We demonstrated that a very good convergence
of the ground-state properties of these nuclei remains unchanged
when the NNN interaction is added to the Hamiltonian. Expressions for the
local $\chi$EFT NNN interaction matrix elements derived in this paper may
by used after some modifications with other bases, e.g. with the hyperspherical 
harmonics basis. 
%
\begin{acknowledgments}
I would like to thank U. van Kolck, E. Epelbaum and J. Adam, Jr. for useful 
comments and A. Nogga for code benchmarking.
This work was performed under the auspices of the
U. S. Department of Energy by the University of California, Lawrence
Livermore National Laboratory under contract No. W-7405-Eng-48. Support
from the LDRD contract No.~04--ERD--058 and from
U.S. DOE/SC/NP (Work Proposal Number SCW0498) is acknowledged.
This work was also supported in part by the Department of Energy under
Grant DE-FC02-07ER41457.
\end{acknowledgments}


\begin{thebibliography}{10}
\expandafter\ifx\csname bibnamefont\endcsname\relax
   \def\bibnamefont#1{#1}\fi
\expandafter\ifx\csname bibfnamefont\endcsname\relax
   \def\bibfnamefont#1{#1}\fi
\expandafter\ifx\csname url\endcsname\relax
   \def\url#1{\texttt{#1}}\fi
\expandafter\ifx\csname urlprefix\endcsname\relax\def\urlprefix{URL }\fi
\providecommand{\bibinfo}[2]{#2}
\providecommand{\eprint}[2][]{\url{#2}}

\bibitem{Weinberg}
\bibinfo{author}{\bibnamefont{{S. Weinberg}}},
  \bibinfo{journal}{Physica} \textbf{\bibinfo{volume}{96A}},
   \bibinfo{pages}{327} (\bibinfo{year}{1979});
  \bibinfo{journal}{Phys. Lett. B} \textbf{\bibinfo{volume}{251}},
   \bibinfo{pages}{288} (\bibinfo{year}{1990});
  \bibinfo{journal}{Nucl. Phys.} \textbf{\bibinfo{volume}{B363}},
   \bibinfo{pages}{3} (\bibinfo{year}{1991});
	J. Gasser {\it et al.}, Ann. of Phys. {\bf 158}, 142 (1984).
        Nucl. Phys. {\bf B250}, 465 (1985).

\bibitem{bernard95}
\bibinfo{author}{\bibnamefont{{V. Bernard, N. Kaiser, and Ulf-G. Mei{\ss}ner
}}},
 \bibinfo{journal}{Int. J. Mod. Phys. E}\textbf{\bibinfo{volume}{4}},
   \bibinfo{pages}{193} (\bibinfo{year}{1995}).

\bibitem{ORK94}
\bibinfo{author}{\bibnamefont{C. Ordonez, L. Ray, and U. van Kolck}},
  \bibinfo{journal}{Phys. Rev. Lett.} \textbf{\bibinfo{volume}{72}},
   \bibinfo{pages}{1982} (\bibinfo{year}{1994}).
  \bibinfo{journal}{Phys. Rev. C} \textbf{\bibinfo{volume}{53}},
   \bibinfo{pages}{2086} (\bibinfo{year}{1996}).

\bibitem{Bira}
\bibinfo{author}{\bibnamefont{{U. van Kolck}}},
  \bibinfo{journal}{Prog. Part. Nucl. Phys.} \textbf{\bibinfo{volume}{43}},
   \bibinfo{pages}{337} (\bibinfo{year}{1999}).

\bibitem{bedaque02a}
\bibinfo{author}{\bibnamefont{P. F. Bedaque and U. van Kolck}},
 \bibinfo{journal}{Ann. Rev. Nucl. Part. Sci. }\textbf{\bibinfo{volume}{52}},
   \bibinfo{pages}{339} (\bibinfo{year}{2002});
   E. Epelbaum, Prog. Part. Nucl. Phys. {\bf 57}, 654 (2006).

\bibitem{Beane06} S. R. Beane, P. F. Bedaque, K. Orginos, and M. J. Savage,     
        Phys. Rev. Lett. {\bf 97}, 012001 (2006). 

\bibitem{NTK05} S. R. Beane, P. F. Bedaque, M. J. Savage and U. van Kolck,
  Nucl. Phys. {\bf A700}, 377 (2002); 
A. Nogga, R. G. Timmermans, and U. van Kolck, 
                Phys. Rev. C {\bf 72}, 054006 (2005).

\bibitem{Birse} M. C. Birse, Phys. Rev. C {\bf 74}, 014003 (2006).

\bibitem{EM06} E. Epelbaum and U.-G. Meissner, nucl-th/0609037.

\bibitem{vanKolck:1994} \bibinfo{author}{\bibnamefont{{U. van Kolck}}},
  \bibinfo{journal}{Phys. Rev. C} \textbf{\bibinfo{volume}{49}},
   \bibinfo{pages}{2932} (\bibinfo{year}{1994}).

\bibitem{Epelbaum:2002}
\bibinfo{author}{\bibfnamefont{E. Epelbaum, 
   A. Nogga, W. Gl\"ockle, H. Kamada, Ulf-G. Meissner and H. Witala}},
   \bibinfo{journal}{Phys. Rev. C} \textbf{\bibinfo{volume}{66}},
   \bibinfo{pages} {064001} (\bibinfo{year}{2002}).

\bibitem{Epelbaum06} E. Epelbaum, Phys. Lett. B {\bf 639}, 456 (2006).

\bibitem{N3LO}
\bibinfo{author}{\bibnamefont{{D. R. Entem and R. Machleidt}}},
  \bibinfo{journal}{Phys. Rev. C} \textbf{\bibinfo{volume}{68}},
   \bibinfo{pages}{041001(R)} (\bibinfo{year}{2003}).

\bibitem{Navratil:2007} P. Navr{\'a}til and V. G. Gueorguiev and J. P. Vary, 
                        W. E. Ormand and A. Nogga, Phys. Rev. Lett. {\bf 99}, 042501 (2007);
                        nucl-th/0701038.

\bibitem{NCSMC12} P. Navr\'atil, J. P. Vary and B. R. Barrett,
                   Phys. Rev. Lett. {\bf 84}, 5728 (2000);
                   Phys. Rev. C {\bf 62}, 054311 (2000).
\bibitem{NO03}
\bibinfo{author}{\bibfnamefont{P. Navr\'atil and W.~E. Ormand}},
   \bibinfo{journal}{Phys. Rev. C} \textbf{\bibinfo{volume}{68}},
   \bibinfo{pages}{034305} (\bibinfo{year}{2003}).

\bibitem{Quaglioni:2007} S. Quaglioni and P. Navr\'atil, Phys. Lett. B (2007), 
                         doi:10.1016/j.physletb.2007.06.082;
                         arXiv:0704.1336.

\bibitem{Nogga06}
\bibinfo{author}{\bibfnamefont{A.~Nogga, P.~Navr\'atil, B.~R. ~Barrett 
   and J.~P. ~Vary}} 
\bibinfo{journal}{Phys. Rev. C} \textbf{\bibinfo{volume}{73}},
\bibinfo{pages}{064002} (\bibinfo{year}{2006}).

\bibitem{UIX}
\bibinfo{author}{\bibfnamefont{B. S. Pudliner, V. R. Pandharipande, J. Carlson,
               and R. B. Wiringa}},
  \bibinfo{journal}{Phys. Rev. Lett.} \textbf{\bibinfo{volume}{74}},
  \bibinfo{pages}{4396} (\bibinfo{year}{1995}).


\bibitem{TM}
\bibinfo{author}{\bibnamefont{{S. A. Coon, M. D. Scadron, P. C. McNamee, B. R.
  Barrett, D. W. E. Blatt and B. H. J. McKellar}}}, \bibinfo{journal}{Nucl.
  Phys. A} \textbf{\bibinfo{volume}{317}}, \bibinfo{pages}{242}
  (\bibinfo{year}{1979}).

\bibitem{TMp} J. L. Friar, D. H\"uber, and U. van Kolck, Phys. Rev. C {\bf 59}, 53 (1999).

\bibitem{TMprime99}
\bibinfo{author}{\bibnamefont{{S. A. Coon}}} \bibnamefont{and}
  \bibinfo{author}{\bibnamefont{{H. K. Han}}}, \bibinfo{journal}{Few-Body
  Systems} \textbf{\bibinfo{volume}{30}}, \bibinfo{pages}{131}
  (\bibinfo{year}{2001}).

\bibitem{Pieper:04} S. C. Pieper, Nucl. Phys. {\bf A571}, 516 (2005).

\bibitem{GFMC_exc_6_8} S. C. Pieper, R. B. Wiringa and J. Carlson,
         Phys. Rev. C {\bf 70}, 054325 (2004).

\bibitem{AV18} R. B. Wiringa, V. G. J. Stoks, and R. Schiavilla,
         Phys. Rev. C {\bf 51}, 38 (1995).

\bibitem{GFMC_IL} S. C. Pieper, V. R. Pandharipande, R. B. Wiringa, and J. Carlson,
         Phys. Rev. C {\bf 64}, 014001 (2001).

\bibitem{GFMC_9_10}
\bibinfo{author}{\bibnamefont{{S. C. Pieper, K. Varga and R. B.
  Wiringa}}}, \bibinfo{journal}{Phys. Rev. C}
  \textbf{\bibinfo{volume}{66}}, \bibinfo{pages}{044310}
  (\bibinfo{year}{2002});
\bibinfo{author}{\bibnamefont{{R. B. Wiringa and S. C. Pieper}}},
\bibinfo{journal}{Phys. Rev. Lett.}
\textbf{\bibinfo{volume}{89}}, \bibinfo{pages}{182501}
  (\bibinfo{year}{2002}).

\bibitem{CG81} S. A. Coon and W. Gl\"ockle, Phys. Rev. C {\bf 23} 1790, (1981).

\bibitem{Friar88} J. L. Friar, B. F. Gibson, G. L. Payne, and S. A. Coon,
               Few-Body Systems {\bf 5}, 13 (1988).

\bibitem{CP93} S. A. Coon and M. T. Pe\~na, Phys. Rev. C {\bf 48}, 2559 (1993).

\bibitem{Huber97} D. H\"uber, H. Witala, A. Nogga, W. Gl\"ockle, and H. Kamada,
               Few-Body Systems {\bf 22}, 107 (1997).

\bibitem{Huber01} D. H\"uber, J. L. Friar, A. Nogga, H. Witala, and U. van Kolck,
               Few-Body Systems {\bf 30}, 95 (2001).

\bibitem{Barnea04} N. Barnea, V. D. Efros, W. Leidemann, and G. Orlandini,
               Few-Body Systems {\bf 35}, 155 (2004).

\bibitem{Adam04} J. Adam, Jr., M. T. Pe\~na, and A. Stadler,
                Phys. Rev. C {\bf 69}, 034008 (2004).

\bibitem{TIHO:2000} P. Navr\'atil, G. P. Kamuntavi\v{c}ius, and B. R. Barrett, 
                   Phys. Rev. C {\bf 61}, 044001 (2000).

\bibitem{NCSM_TM} D. C. J. Marsden, P. Navr\'atil, S. A. Coon and B. R. Barrett,
                  Phys. Rev. C {\bf 66}, 044007 (2002).

\bibitem{Nogga_pc} A. Nogga, private communication.

\bibitem{Bira_cD}  C. Hanhart, U. van Kolck, and G. A. Miller,
                  Phys. Rev. Lett. {\bf85}, 2905 (2000).

\bibitem{NO02} P.~Navr{\'a}til and W.~E. Ormand, 
               Phys. Rev. Lett. {\bf 88} (2002) 152502.

\bibitem{NC04} P.~Navr{\'a}til and E.~Caurier, Phys. Rev. C {\bf 69} (2004) 014311.

\bibitem{HHnonloc} M. Viviani, L. E. Marcucci, S. Rosati, A. Kievsky 
and L. Girlanda, Few-Body Systems {\bf 39}, 159 (2006).


\end{thebibliography}
\end{document}